\documentclass[preprint,12pt,times]{elsarticle}
\usepackage[T1]{fontenc}
\usepackage{graphicx}
\usepackage{epstopdf}
\usepackage{amssymb}
\usepackage{amsfonts,amsmath,bm,bbm}
\usepackage{mathtools}
\usepackage{enumerate}
\usepackage{indentfirst}
\usepackage[font=footnotesize]{caption}
\usepackage{floatrow}
\usepackage{enumitem}
\usepackage[table,xcdraw,dvipsnames]{xcolor}
\usepackage{url}
\usepackage{multirow}
\usepackage{multicol}
\usepackage{bbm}

\usepackage{lineno}

\usepackage[a4paper,
 left=2cm,
 right=2cm,
 ]{geometry}
 
\usepackage{enumitem}
\DeclareMathOperator{\argmin}{argmin}

\definecolor{dcyan}{rgb}{0.1,0.3,0.6}

\makeatletter
\def\ps@pprintTitle{%
 \let\@oddhead\@empty
 \let\@evenhead\@empty
 \let\@oddfoot\@empty
 \let\@evenfoot\@oddfoot
}
\makeatother

\journal{Renewable Energy}

\begin{document}

\begin{frontmatter}

\title{The role of probabilistic load and renewable prediction in enhancing day-ahead electricity price forecasts}

\author[inst1]{Bartosz Uniejewski\corref{cor1}}
\ead{bartosz.uniejewski@pwr.edu.pl}

\affiliation[inst1]{organization={Department of Operations Research and Business Intelligence},
 addressline={Wroc{\l}aw University of Science and Technology}, 
 city={Wroc{\l}aw},
 postcode={50-370},
 country={Poland}}

\author[inst2]{Florian Ziel}

\affiliation[inst2]{organization={House of Energy, Climate and Finance},
 addressline={University of Duisburg-Essen}, 
 city={Essen},
 postcode={45141},
 country={Germany}}

\cortext[cor1]{Corresponding author}

\begin{abstract}
The increasing penetration of renewable energy sources (RES) has amplified volatility in electricity markets, while demand variability continues to challenge system stability. Traditional day-ahead electricity price forecasting (EPF) models rely on point forecasts of load and RES generation, which fail to capture the uncertainty inherent in weather-driven supply and dynamic demand. This study introduces a novel framework that integrates probabilistic forecasts of load, wind, and solar generation into EPF models. Using data from the German EPEX market between 2015 and 2023, we generate quantile forecasts of fundamental variables via
{historical simulation, conformal prediction and quantile regression}, and incorporate them into both parsimonious expert and high-dimensional LASSO-based models. Empirical results show that probabilistic inputs substantially enhance forecast accuracy, reducing root mean square errors by up to 13\% compared with point-forecast benchmarks. Notably, extreme quantiles of load and RES forecasts emerge as the most influential predictors, underscoring the importance of rare but system-critical scenarios. These findings demonstrate that accounting for uncertainty in both load and renewables is crucial for reliable electricity price forecasting and offers practical value for system operators, traders, and policymakers navigating renewable integration.
\end{abstract}


\end{frontmatter}

\section{Introduction and motivation}

Recently the energy sector has become more dynamic as renewable energy sources (RES) have expanded and integrated into existing infrastructure in order to meet the European Union's 2050 carbon neutrality target. In 2023, 45\% of Germany's electricity needs were covered by renewables (31\% wind, 12\% solar, and 2\% other RES). However, wind and solar generation depend on weather conditions. For example, on some days, RES covers 95\% of national electricity demand (i.e. 29.12.2023), whereas on others the production drops below 7\% (i.e. 30.11.2023). 

This unpredictability challenges system operators and increases price volatility, complicating decisions for market participants \citep{kab:etal:24}. Furthermore, market participants face additional challenges as electricity systems become increasingly linked with other sectors, such as transport through electric vehicles \citep{qin:etal:24} and heating through combined heat and power (CHP) systems \citep{had:etal:26}. This makes energy forecasting increasingly important in the coming decades \cite{lag:mar:des:wer:21,pol:chy:21}.

Instead of expanding infrastructure, utilities can try to slow the growth of peak demand through demand-side management (DSM). This involves changing end-users’ consumption patterns through regulation, financial incentives, or social pressure \citep{sko:etal:24}. However, in combination with the variability of RES, DSM and sector coupling increase the unpredictability of net load. As a result, sudden drops in both generation and consumption become more likely, amplifying imbalances and price volatility \citep{gia:par:pel:16,hon:pin:etal:20}. Under such conditions, forecasting electricity prices (EPF) becomes challenging \citep{lag:mar:des:wer:21}, and models built a few years ago no longer provide accurate predictions. This situation calls for innovative forecasting methods that surpass current methods and meet today's electricity markets' demanding requirements.

Recent empirical findings indicate that incorporating forecasts of fundamental variables, such as electricity demand and solar and wind generation, into EPF models significantly improves performance compared to models without these variables \citep{hui:ste:22}.
However, to the best of our knowledge, all existing approaches rely exclusively on point forecasts. This overlooks the fact that load and RES are inherently uncertain. Ignoring this issue is problematic because electricity prices are determined by a nonlinear supply-demand equilibrium in which the distribution, not just the expected value, of fundamentals shapes market outcomes.

\begin{figure}[htb!]
 \centering
 \includegraphics[width=.7\textwidth]{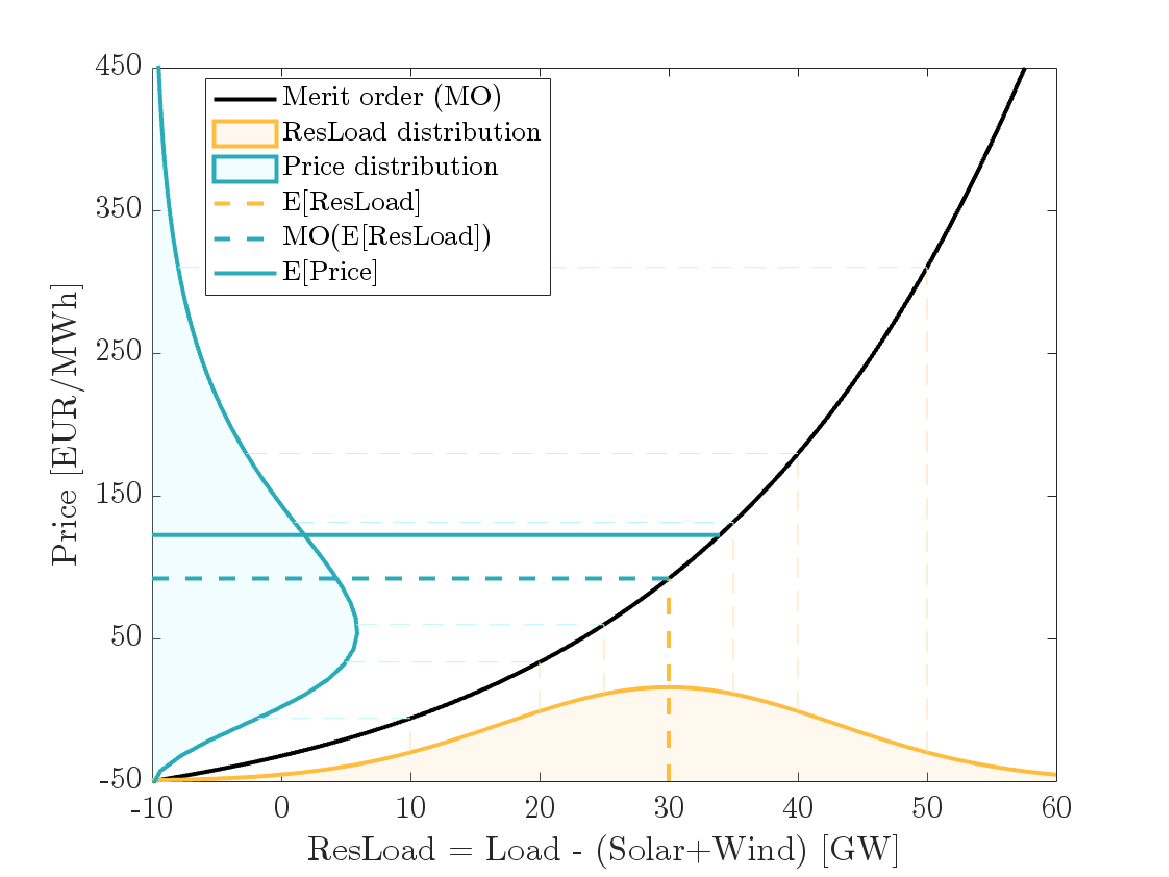}
 \caption{
 Illustrative supply-stack model with merit order curve (black) depending on the residual load (orange) and resulting electricity price (cyan).}
 \label{fig:motivation}
\end{figure}

In this study, we argue that electricity price forecasting models should include not only point forecasts of fundamental variables, but also probabilistic forecasts.
The relevance of probabilistic inputs in these models is well justified by fundamental theory. Electricity prices result from the intersection of supply and demand curves, which are typically non-linear. Figure~\ref{fig:motivation} shows an example of a supply stack model with a nonlinear merit order (MO) curve that depends on the residual load (ResLoad), which is the total load minus solar and wind generation. In a deterministic setting, a ResLoad of 30 GW leads to an electricity price of MO(ResLoad)=92 EUR/MWh. However, if the residual load is uncertain, as it is in day-ahead markets, with a density centered around 30 GW, as shown in Figure~\ref{fig:motivation}, then the electricity price also becomes uncertain, or a random variable, as shown by the cyan density on the y-axis. However, the expected value of this cyan electricity price density is approximately $\text{E}[\text{MO}(\text{ResLoad})] = 122.44 \text{EUR/MWh}$. This is significantly higher than the point forecast plugged into the merit order:
$\text{MO}(\text{E}[\text{ResLoad}])=92\text{EUR/MWh}$. Thus, to properly derive the expected electricity price, the full distribution of residual loads or, more generally, the distribution of electricity loads such as solar and wind generation is required. Therefore, our fundamental theory leads us to use probabilistic forecasts of load and renewable energy sources (RES) to derive accurate point forecasts of electricity prices.

Current forecasting frameworks fail to leverage the rich information contained in probabilistic forecasts, which can better reflect nonlinear effects—such as merit order curve convexity—on price outcomes. To date, no study has systematically evaluated the impact of using full quantile distributions of fundamental variables as exogenous inputs in electricity price models. This paper addresses this critical gap by designing a model that includes -- in addition to the standard set of regressors -- a set of quantile forecasts of fundamental variables. To estimate the models, we use the Least Absolute Shrinkage and Selection Operator \citep[LASSO; ][]{tib:96}, which allows us to consider an almost unlimited number of explanatory variables. Our empirical tests, performed on the German electricity market, show that the integrating of probabilistic inputs significantly improves the accuracy of electricity price forecasts. The proposed methods outperform the standard approach by up to 13.3 \%.

{The remainder of this paper is structured as follows. In Section \ref{sec:postprocessing} we explain how the probabilistic forecasts of fundamental variables are computed and then in Section \ref{sec:methodology} we show how they are included in the the day-ahead electricity prices forecasting models. In Section \ref{sec:Results} we start with presenting the datasets, than we discuss the accuracy probabilistic forecasts of fundamental variables, finally we present and compare obtained electricity price forecasts in terms of the linear and quadratic measures and the test for Conditional Predictive Ability (CPA) of Giacomini and White \citep{gia:whi:06}. We conclude Section \ref{sec:Results} with the analysis of the most frequently selected variables. Finally, in Section \ref{sec:conclusion} we summarize the main results.}

\section{Postprocessing of point predictions of fundamental variables}
\label{sec:postprocessing}
Publicly available data provides point predictions of day-ahead load and forecasts of solar and wind generation. Since these variables are the key sources of uncertainty in RES-dominated power systems, we transform them into probabilistic forecasts to better reflect their variability. This allows the forecasting models to explicitly account for uncertainty in demand and renewable generation rather than relying on single deterministic values.
To obtain the probabilistic forecasts we utilize a post-processing technique that converts point forecasts into quantile forecasts \citep{lip:uni:wer:24}. Although there are several well-established post-processing techniques, we chose to compare two approaches: historical simulation and quantile regression.

Before describing the methods, we would like to point out that, due to the significant seasonality of solar power generation, which can lead to inaccurate estimates, we have chosen not to construct a probabilistic forecast when the day-ahead point forecast is zero ($S_{d,h} = 0$). Additionally, the post-processing methods do not impose any restrictions on the quantile forecast being non-negative. Therefore, for variables that theoretically cannot be negative, we truncated them to zero.

\begin{figure}[htb!]
 \includegraphics[width = 0.6\linewidth]{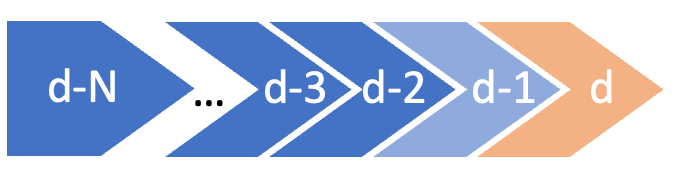}
 \caption{Timeline with data availability at the time of the forecast. The forecast for day $d$ (in orange) is made on day $d-1$. From day $d-N$ to day $d-2$ (in dark blue), we have access to both the day-ahead forecast and the actual observations. The situation changes on the day of the forecast ($d-1$, marked in light blue), on this day we already have access to the day-ahead forecast for days $d-1$ and $d$, but the corresponding actual observations are still missing.}
 \label{fig:prob}
\end{figure}

In addition, we would like to point out that both historical simulation and quantile regression require access to actual observations of the fundamental variable of interest.
It is important to keep in mind that at the beginning of day $d-1$ (at the moment of forecasting day $d$), we only have access to actual observations up to day $d-2$ (see Figure~\ref{fig:prob}). Taking this into account, we calibrate the model and obtain the empirical quantile of the errors ($\hat q^{\text{emp}}_\tau$ {for HS and CP in eq. \refeq{eq:HS} and \refeq{eq:CP})} or coefficients $\beta_0^{\tau}$ and $\beta_1^{\tau}$ (for QR in eq. \refeq{eq:QR}) we use data from $d-N$ to $d-2$ ($N$ in the length of the calibration window size, here $N=364$).

Finally, we note that we applied the proposed post-processing techniques to obtain probabilistic forecasts of system load, solar and wind power generation, and a linear combination of these: RES (solar + wind) and ResLoad (load - RES). These variables capture the most important drivers of short-term price volatility in electricity markets, particularly under high renewable penetration. Although RES and ResLoad are defined as linear combinations, their probabilistic forecasts are not constructed by summing the quantiles of their component forecasts. Rather, RES and ResLoad are treated as independent variables with their own empirical error distributions or quantile regressions based on their joint historical behavior. This approach implicitly captures dependency patterns between components without relying on convolution-based methods.

\subsection{Probabilistic forecasting by historical simulation}
The historical simulation (HS) method is a straightforward approach. We begin by computing the empirical quantile $q_\tau^{\text{emp}}(\epsilon_{d,h})$ of the point prediction error $\epsilon_{d,h} = x_{d,h} - \hat{x}_{d,h}$. Note that the empirical quantile is estimated using the historical distribution of prediction errors $\epsilon_{d*,h} = x_{d*,h} - \hat{x}_{d*,h}$ for ${d* = d-N, \ldots, d-2}$, where $N$ is the size of the calibration window (in our case, $N = 364$). This value is then added to the point prediction to obtain the quantile prediction of $x_{d,h}$ for day $d$ and hour $h$:

\begin{equation}
\label{eq:HS}
	\hat{q}_\tau^{\text{HS}}({x}_{d, h}) = \hat{x}_{d, h}+\hat{q}_\tau^{\text{emp}}(\epsilon_{d,h}), 
\end{equation}

where $\tau$ is a probability level, $x_{d,h}$ is the actual observation of given fundamental variable for day $d$ and hour $h$ and $\hat x_{d,h}$ is corresponding day-ahead point forecast.

\subsection{{Probabilistic forecasting by conformal prediction}}

{Conformal prediction provides a framework for constructing prediction intervals with finite-sample validity guarantees \citep{lei2018distribution}. In contrast to historical simulation, which directly uses signed forecast errors, conformal prediction employs absolute errors to create symmetric intervals centered on the point forecast.

We begin by calculating absolute forecast errors on the calibration window of $N=364$ days:
\begin{equation}
e_{d*,h} = |x_{d*,h} - \hat{x}_{d*,h}|, 
\end{equation}
for $d* = d-N, \ldots, d-2$.
For a target quantile level $\tau \in (0,1)$, we construct prediction intervals by computing the empirical quantile of absolute errors. Specifically, for lower quantiles ($\tau < 0.5$), we subtract the corresponding error quantile from the point forecast; for upper quantiles ($\tau > 0.5$), we add it. This yields:

\begin{equation}
\label{eq:CP}
\hat{q}_\tau^{\text{CP}}({x}_{d, h}) = \hat{x}_{d,h} + \text{sign}(\tau - 0.5) \cdot \hat{q}_{|2\tau-1|}^{\text{emp}}(e_{d,h}),
\end{equation}

where $\text{sign}(\cdot)$ denotes the sign function and $\hat{q}_{|2\tau-1|}^{\text{emp}}(e_{d,h})$ is the empirical quantile of the absolute errors at level $|2\tau-1|$. For instance, to construct the 5th and 95th percentiles, we compute the 90th percentile of absolute errors ($\hat{q}_{0.9}^{\text{emp}}(e_{d,h})$) and form the interval $[\hat{x}_{d,h} - \hat{q}_{0.9}^{\text{emp}}, \hat{x}_{d,h} + \hat{q}_{0.9}^{\text{emp}}]$.
This approach imposes symmetry on the prediction intervals, which can be beneficial when computational simplicity is desired or when there is no strong evidence of directional bias in forecast errors.}

\subsection{Probabilistic forecasting by quantile regression}
Quantile regression (QR) is a method that has been used successfully in many forecasting applications:
in load forecasting
\cite{liu:now:hon:wer:17,wan:etal:19}, in electricity price forecasting \cite{cor:din:pou:24} and also renewable energy forecasting \cite{hu2022conformalized, yan:yan:liu:23}. This approach estimates conditional quantiles of the target variable as a linear combination of point forecasts in a quantile regression setting \citep{koe:05}:
\begin{equation}
\label{eq:QR}
 \hat{q}_\tau^{\text{QR}}({x}_{d, h}) 
=
 \beta_0^{\tau} + \beta_1^{\tau}\hat{x}_{d, h}, 
\end{equation}
where $\hat{q}_\tau^{\text{QR}}$ is the conditional $\tau$th quantile and $\beta_i^{\tau}$ are coefficients that are estimated independently for each quantile and each hour of a day by minimizing so-called check function:

\begin{equation}\label{eq:qr:loss}
 \argmin_{\beta_0,\beta_1}\sum_{d = 1}^N \text{QL}_\tau(\beta_0 + \beta_1\hat{x}_{d, h},{x}_{d,h})
\end{equation}
with quantile loss $\text{QL}_\tau(q,x) = (\mathbbm{1}_{\{x<q\}}-\tau)(q-x)$.

Note that the numerical inefficiency of quantile regression can result in overlapping neighboring percentiles, leading to a quantile crossing problem \citep{che:f-v:gal:10}. To address this, following \cite{mac:now:16} the quantiles are sorted to obtain monotonic quantile curves, independently for each day and hour.

\subsection{ReLU-based transformation}
Furthermore, for benchmarking purposes, we apply nonlinear transformations of the $hat{x}_{d,h}$ forecasts. To ensure that the improvement in forecast accuracy is due to the use of probabilistic forecasts and not simply the introduction of a nonlinearity in the model, we apply the following transformation:

\begin{equation}
\hat{q}_\tau^{\text{ReLU}} ({x}_{d, h}) = \max\{\hat{x}_{d, h}, \hat{q}_{\tau}^{\text{emp}}(\hat x_{d,h}) \} = 
\text{ReLU}( \hat{x}_{d, h} - \hat{q}_{\tau}^{\text{emp}}(\hat x_{d,h}) ) + \hat{q}_{\tau}^{\text{emp}}(\hat x_{d,h}) 
\end{equation}
where $\hat{q}_{\tau}^{\text{emp}}(\hat x_{d,h})$ is an empirical $\tau$-quantile of day-ahead forecast of given fundamental variable from the calibration window.

\begin{figure*}[htb!]
 \includegraphics[width = 0.99\linewidth]{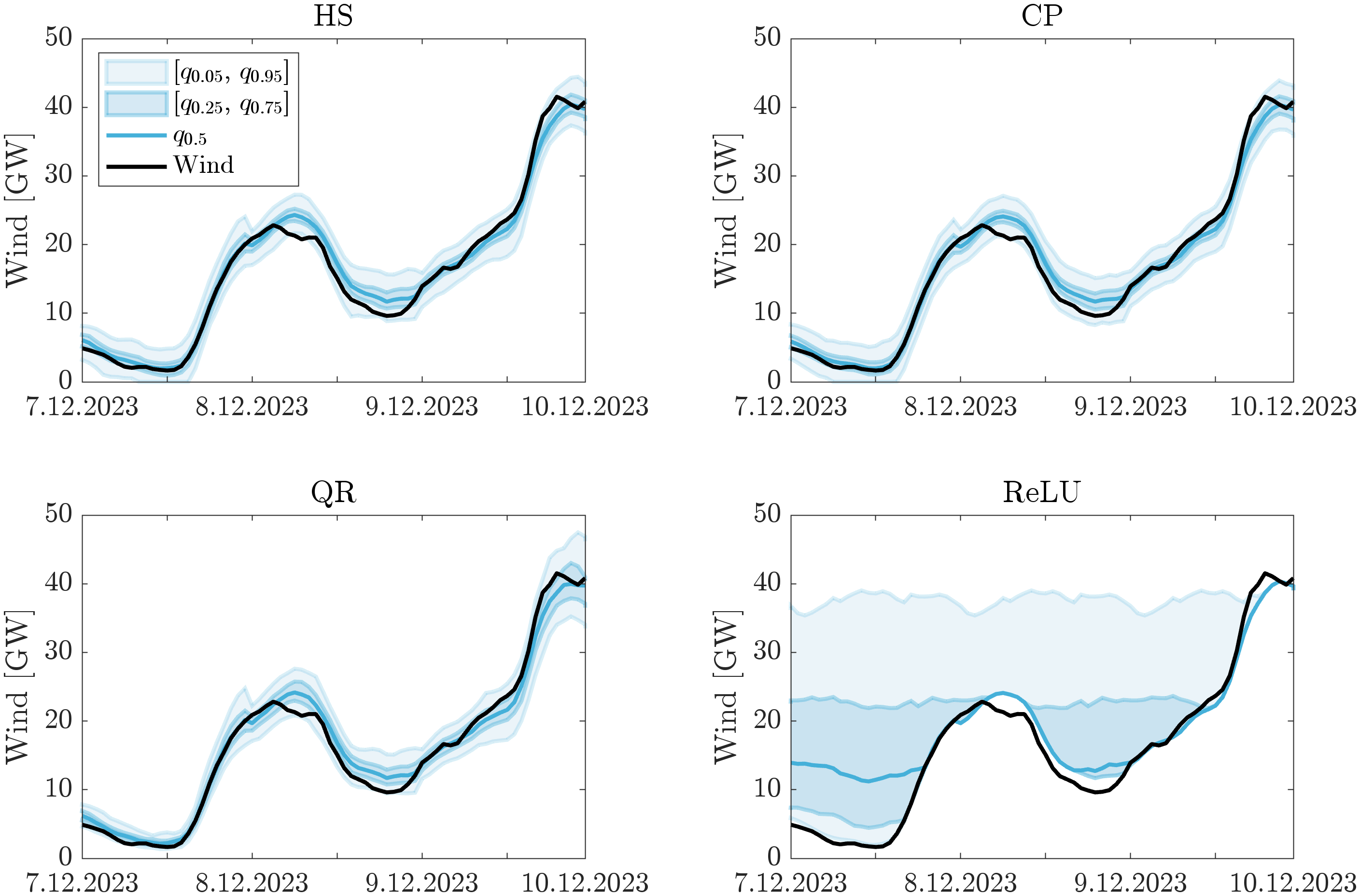}
 \caption{ Comparison of predictive distribution of wind power generation for days from 7.12.2023 to 9.12.2023 for historical simulation (upper left), conformal prediction (upper right), quantile regression (lower left) and ReLU-based transformation (lower right).}
 \label{fig:prob_comparison}
\end{figure*}

Note that unlike the other two approaches, the ReLU-based method does not produce the quantile prediction of $\hat{x}_{d, h}$ (see Figure~\ref{fig:prob_comparison}). Instead, it produces a nonlinear transformation of the empirical quantiles of $\hat{x}_{d, h}$. This means that unlike $\hat{q}_\tau^{\text{HS}}$ and $\hat{q}_\tau^{\text{QR}}$, the expression $\hat{q}_\tau^{\text{ReLU}}$ is not the quantile prediction. Keep in mind that the value of $\hat{q}_\tau^{\text{ReLU}}$ is calculated using only the historical data on the forecasts of $\hat{x}_{d,h}$, without the necessity of an actual observation of $x_{d,h}$, so it is derived using the entire calibration window from day $d-N$ to $d-1$. 


\section{Methodology}
\label{sec:methodology}

\subsection{Base Models}

\subsubsection{Expert model}
To assess the impact of uncertainty in load and renewable generation, we embed probabilistic forecasts into two established model classes for day-ahead price forecasting.
 The first is a parsimonious linear model which contains autoregressive information with exogenous variables (later referred to as the Expert model). Various versions of this model were used in a number of EPF studies under different names and abbreviations \citep[see, for example:][]{bil:gia:del:rav:23,hub:mar:wer:19}. The electricity price for day $d$ and hour $h$ is given by the following formula: 

\begin{align}
 p_{d,h} & = \beta_0 + \beta_1 p_{d-1,h} + \beta_2 p_{d-2,h} + \beta_3 p_{d-7,h} + \beta_4 p_{d-1,24} + \beta_5 p_{d-1}^{\min} + \beta_6 p_{d-1}^{\max} \nonumber\\ 
 & + \beta_7 \widehat{\text{Load}}_{d,h} + \beta_8 \widehat{\text{Solar}}_{d,h} + \beta_9 \widehat{\text{Wind}}_{d,h} 
+ \beta_{10} \text{Coal}_{d-2}^{\text{close}} + \beta_{11} \text{Gas}_{d-2}^{\text{close}} \nonumber\\ & + \beta_{12} \text{Oil}_{d-2}^{\text{close}}+ \beta_{13} \text{EUA}_{d-2}^{\text{close}}
 + \sum_{i=1}^7 \beta_{13+i} D_i + \varepsilon_{d,h}
 \label{eq:Expert}
\end{align}
where the first three regressors capture the autoregressive effects of the prices for the same hour on days $d-1$, $d-2$ and $d-7$, $p_{d-1,24}$ provides information on the last known price, i.e, midnight of day $d-1$, $p^{\max}_{d-1}$ and $p^{\min}_{d-1}$ stand for the maximum and minimum price of the previous day, $\widehat {\text{Load}}_{d,h}$, $\widehat {\text{Solar}}_{d,h}$ and $\widehat {\text{Wind}}_{d,h}$ stand for the day-ahead forecasts of system-wide load, solar and wind generation, respectively. The next four variables represent the last observed market prices (i.e., the closing price of day $d-2$) for coal, gas, oil, and carbon emission allowances (EUAs). Finally, $D_1,\dots,D_7$ are daily dummies.

\subsubsection{High-dimensional linear model}

The second model class is a parameter-rich high-dimensional linear model (HLM), which has been used and modified in several EPF studies \citep{wag:ram:sch:mic:22, hirsch2025online}.
As this model type is often (also here) estimated using LASSO, literature calls this LEAR (LASSO Estimated AutoRegressive model)\footnote{Since both Expert and HLM models are estimated using LASSO, we want to avoid the LEAR notation.}.

\begin{align}
 p_{d,h} & = \beta_0 + \sum_{i=1}^{24} \left( \beta_i p_{d-1,i} + \beta_{24+i} p_{d-7,i} \right) + \beta_{49} p_{d-1}^{\min} + \beta_{50} p_{d-1}^{\max} + \sum_{i=1}^{24} \left( \beta_{50+i} \widehat{\text{Load}}_{d,i} + \beta_{74+i} \widehat{\text{Load}}_{d-1,i} \right) \nonumber\\
 & + \sum_{i=1}^{24} \left(\beta_{98+i} \widehat{\text{Solar}}_{d,i} +\beta_{122+i} \widehat{\text{Solar}}_{d-1,i}\right) 
 + \sum_{i=1}^{24} \left(\beta_{146+i} \widehat{\text{Wind}}_{d,i} + \beta_{170+i} \widehat{\text{Wind}}_{d-1,i}\right) 
 \nonumber\\
 & + \beta_{195} \text{Coal}_{d-2}^{\text{close}}+ \beta_{196} \text{Gas}_{d-2}^{\text{close}} + \beta_{197} \text{Oil}_{d-2}^{\text{close}} + \beta_{198} \text{EUA}_{d-2}^{\text{close}}
 + \sum_{i=1}^7 \beta_{198+i} D_i +  \varepsilon_{d,h}
 \label{eq:HLM}
\end{align}

The over 200-variable structure of this model allows all cross-hour dependencies to be included in the price forecasts. 

Note that to avoid overfitting, the $\widehat{\text{Solar}}_{d,h}$ are excluded from for both Expert and HLM models for hours where more than 25\% of the observations in the calibration window are equal to 0.

\subsection{Proposed extensions}

In this study, we propose to include in forecasting models additional information that addresses the uncertainty of load, wind, and solar generation. The idea is to incorporate quantile forecasting of fundamental variables into the model structure:

\begin{equation}
 p_{d,h} = \text{BaseModel} + \sum_{x \in \mathcal{X}} \sum_{\tau \in \mathcal{T}} \beta_{x,\tau} q^*_\tau(x_{d,h}),
 \label{eq:extension}
\end{equation}

where $\text{BaseModel}$ indicates the model structure of either an Expert (Eq. \eqref{eq:Expert}) or HLM-type (Eq. \eqref{eq:HLM}) model, $\mathcal{X}$ is a set of fundamental drivers (load and/or renewable generation) included in the model, $\mathcal{T}$ is a set of considered probability levels and $*$ refers to one of the postprocessing methods. 

We considered several variants of our approach depending on several factors. Let us now briefly describe all the features that define a given model. First, as already described in Section \ref{sec:postprocessing}, we consider {three} different postprocessing methods to obtain the quantile forecast of the fundamental variables: historical simulation $q^{\text{HS}}(x_{d,h})$, {conformal prediction historical simulation $q^{\text{CP}}(x_{d,h})$} and quantile regression averaging $q^{\text{QR}}(x_{d,h})$. Alternatively, the postprocessing technique can be replaced by a ReLU-based nonlinear transformation to obtain $q^{\text{ReLU}}(x_{d,h})$.

{Second, we consider different sets of fundamental variables (load, solar and wind) to be input into the model. 
To be more specific, we define seven different sets $\mathcal{X}_1,\ldots, \mathcal{X}_7$ for $\mathcal{X}$ motivated by plausible fundamental modeling perspective in electricity markets
that carry some information of Load, Solar and Wind:
}

\begin{itemize}[leftmargin=10mm]
 \item[\textbf{1-3}] {The first three are one-element information sets, where we include in the model the quantile forecast of either \mbox{$\mathcal{X}_1 = \{\text{Load}\}$}, \mbox{$\mathcal{X}_2 = \{\text{Solar}\}$}, or \mbox{$\mathcal{X}_3 = \{\text{Wind}\}$}, but only one at a time.}
 \item[\textbf{4~~}] { Second, we consider another one-element set, but already using the information of two variables, namely \mbox{$\mathcal{X}_4 = \{\text{RES}\}$} where $\text{RES}= \text{Wind}+\text{Solar}$ aggregates all the fluctuating renewable energy production from wind and solar. }
 \item[\textbf{5-7}] { The last three variants include information about all three considered fundamental variables, but with different levels of aggregation. We consider a one-element set including the residual load \mbox{$\mathcal{X}_5 = \{\text{ResLoad}\}$} with $\text{ResLoad}= \text{Load} -\text{Wind}-\text{Solar}$, a two-element set consisting of Load and RES $\mathcal{X}_6 = \{\text{Load}, \text{RES}\}$, and finally a three-element set $\mathcal{X}_7 = \{\text{Load}, \text{Solar}, \text{Wind}\}$. }
\end{itemize}
{
Note that considering $\text{RES}$ is plausible under the assumption that the effect of one additional unit wind power in the electricity system should have the same effect as one additional unit of solar power in the system. With similar arguments, the effect of generating one unit additional renewable power and reducing the demand by one unit should be the same and motivates the definition of $\text{ResLoad}$.}

Next, we consider seven sets of incorporated probability levels represented by $\mathcal{T}$. Specifically, we compare the predictive performance of models that include a different number of quantile predictions for a given fundamental variable:
\begin{multicols}{2}
\begin{itemize}[leftmargin=10mm,itemsep=0pt]
 \item[$\mathcal{T}_5$] $= \{\gamma, 0.1,0.5,0.9, 1-\gamma\}$,
 \item[$\mathcal{T}_7$] $= \{\gamma, 0.1,0.3,0.5,0.7,0.9, 1-\gamma\}$,
 \item[$\mathcal{T}_{11}$] $= \{\gamma, 0.1,0.2, \ldots, 0.9, 1-\gamma\}$,
 \item[$\mathcal{T}_{21}$] $= \{\gamma, 0.05, 0.1, \ldots, 0.95, 1-\gamma\}$,
 \item[$\mathcal{T}_{51}$] $= \{\gamma, 0.02, 0.04, \ldots, 0.98, 1-\gamma\}$,
 \item[$\mathcal{T}_{101}$] $= \{\gamma, 0.01, 0.02, \ldots, 0.99, 1-\gamma\}$,
 \item[$\mathcal{T}_{201}$] $= \{\gamma, 0.005, 0.01, \ldots, 0.995, 1-\gamma\}$,
 \item[]
\end{itemize}
\end{multicols}
where $\gamma = \frac{1}{2N}$ ($N$ is the length of the calibration window; here $N=364$) is the probability level of the extreme quantiles represents the prediction of the minimum. Equivalently, the $1-\gamma$ quantile corresponds to the maximum level forecast of the predicted fundamental variable. Note that the dense grid of 201 probability levels approaches the maximum number that can be estimated with 364 observations in the calibration window. The next step would be 401 and would provide the same prediction for at least some pairs of quantiles.

Lastly, the models differ in base model structure. Probabilistic inputs can be added to either the parsimonious 20-variable Expert model or the parameter-rich HLM model.

To summarize, we are considering a total of {392} different models depending on the following four factors: (i) the postprocessing (or transformation) technique, (ii) the set of fundamental variables included as probabilistic inputs, (iii) the density of the probability level grid, and (iv) the structure of the base model.
To clarify the notation, we denote each model using the same pattern: $\text{BaseModel}-P^\mathcal{X}_\mathcal{T}$, where $B$ denotes the baseline model, $P$ refers to the postprocessing technique, $\mathcal{X}$ is the set of selected variables, and finally $\mathcal{T}$ is the number of quantiles included in the model. 
For example, 
\mbox{Expert-HS$^\text{Solar}_{\mathcal{T}_5}$} is a parsimonious 'Expert' model extended by 5 additional variables referring to the quantile forecasts of solar power generation These forecasts are obtained using historical simulation $\hat q^{\text{HS}}_\tau(\text{Solar}_{d,h})$ for $\tau = \gamma, 0.1, 0.5, 0.9, 1-\gamma$, while 
\mbox{HLM-QR$^{\{\text{Load, Solar, Wind}\}}_{\mathcal{T}_{201}}$} is a HLM-type structure enriched with 201 quantile forecasts of load, solar and wind power generations obtained with quantile regression (it is the largest model considered with $205+201\times3=808$ variables for a single hour $h$).

\subsection{Estimation and forecasting}

In this study, we generate both probabilistic (quantile) forecasts of fundamental variables and point forecasts of electricity prices. We use a multivariate modeling framework that employs a set of 24 interrelated models, one for each hour of the day $d$.

it The estimation of the $\beta_i$ parameters of both Expert and HLM (Eq.\ \eqref{eq:Expert} and \eqref{eq:HLM}) and all proposed extensions of these models (Eq.\ \eqref{eq:extension}) is based on the LASSO method \citep{tib:96}. Although we do not use a variance stabilizing transformation \citep[VST;][]{uni:wer:zie:18} function, all inputs are standardized and the unpenalized intercept is included in the model. 
Following the conclusion of \cite{uni:24:ORD}, we used cross-validation (CV) to select the value of the tuning parameter. Specifically, we used a 7-fold CV with a random split between the training and test periods to select one (the best performing) from the grid of 100 $\lambda$ values.

\section{Results}
\label{sec:Results}

\subsection{Test ground}
\label{sec:data}

In this study, {we perform our empirical tests on} publicly available data from Transparency ENTSO-E (\url{https://transparency.entsoe.eu}) for the German energy market. Specifically we used day-ahead prices for German-Luxemburg bidding zone\footnote{Can be accessed from ENTSO-E: Transmission $\rightarrow$ Day-ahead Prices with code BZN$|$DE-LU.} (the bidding zone also includes Austria but only until 09/30/2018) and the day-ahead forecasts and actual observations of load, solar and wind generation in Germany\footnote{Can be accessed from ENTSO-E with code CTY$|$DE. Note that we aggregate the values for onshore and offshore wind into one time series}. In addition, we used data from Investing.com (\url{https://www.investing.com}) for the closing prices of API2 (coal), Title Transfer Facility (TTF; natural gas), Brent (crude oil) and European Union Allowance (EUA; carbon emissions).

Note that some of the collected data are available in 15-minute resolutions. Therefore, we aggregated them into hourly time series to unify the dataset. We then pre-processed all hourly series to account for time changes to and from Daylight Saving Time. During the transition from Central European Time (CET) to Central European Summer Time (CEST), missing values are replaced by the arithmetic mean of the surrounding hours' observations. Double values that occur during the transition back to CET are replaced by their mean.

\begin{figure*}[htb!]
 \centering
 \includegraphics[width = 0.99\linewidth]{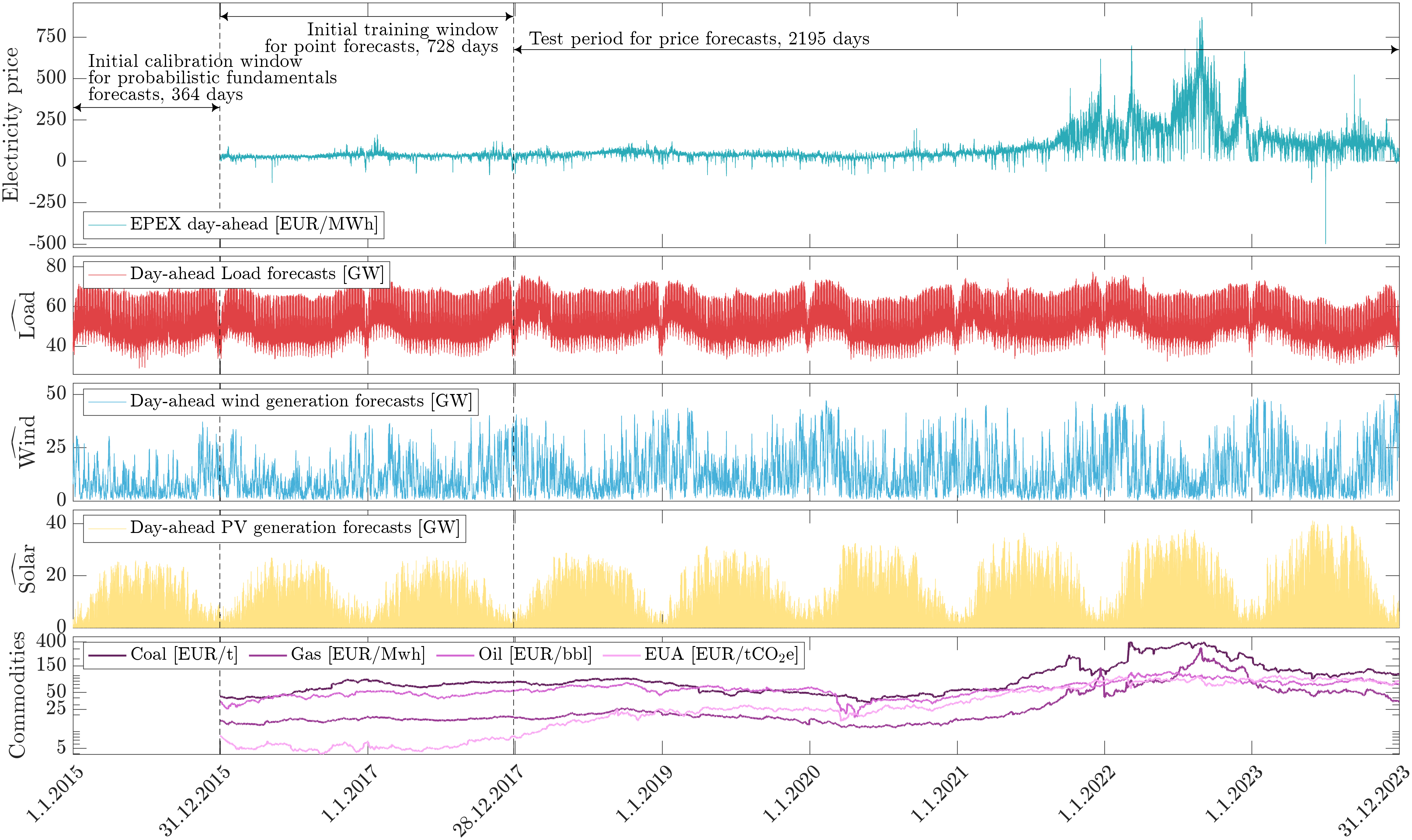}
 \caption{EPEX SPOT hourly day-ahead prices (top), hourly day-ahead forecasts of system load (middle top), wind generation (onshore+offshore; middle) and solar generation (middle bottom) and commodities prices for the period 1.1.2015-31.12.2023. The first vertical dashed line marks the end of the 364-day calibration window for probabilistic forecasting of fundamental variables and the beginning of the 728-day period for training electricity price point forecasting models. The second dashed line marks the beginning of the 2195-day out-of-sample test period. Note that the first 364 days of the price series (both electricity and commodities) are excluded from the plot as they are not used in this study.
 } 
 \label{fig:data}
\end{figure*}

All collected time series span from 1.1.2015 to 31.12.2023, with a 6-year out-of-sample test period starting at 28.12.2017, as shown in Figure~\ref{fig:data}. To obtain the forecast, we use a rolling window scheme. First, we use a 364-day (52 weeks, ca. one year) rolling calibration window to compute the probabilistic forecast of the fundamental variables for the period from 31.12.2015 to 31.12.2023. Once the probabilistic forecast is available, another rolling calibration window of 728 days is introduced to estimate the price forecasting models. For example, to obtain the electricity price forecast for all 24 hours of 29.12.2017, we use the collected time series together with the obtained probabilistic forecast for the period from 31.12.2015 to 28.12.2017, then the calibration window is shifted by one day (1.1.2016-29.12.2017 and the forecast for 30.12.2017 is obtained. The procedure is repeated until the last day of the out-of-sample period is obtained.

\subsection{Accuracy of fundamentals probabilistic forecasts}

{
To ensure transparency and credibility of our framework, we first evaluate the quality of the probabilistic forecasts of fundamental variables that later serve as inputs to the price forecasting models. Although the main contribution of this study lies in demonstrating the value of incorporating such probabilistic inputs into electricity price forecasting, it is important to verify that the underlying forecasts are themselves sufficiently accurate and reliable. This step does not constitute a primary research objective, but rather serves to justify the use of the selected post-processing methods and to confirm that they produce meaningful quantile forecasts of the fundamental variables. 
}

\subsubsection{Evaluation measures}
{
To evaluate the quality of the probabilistic forecasts of load, wind, solar, RES, and residual load (ResLoad), we use the \textit{pinball score} (PB), the standard metric for quantile forecasts. For a quantile level $\tau \in (0,1)$, the score is defined as
\begin{equation}
\text{PB}_\tau(y, \hat{q}_\tau) = \big(\tau - \mathbbm{1}\{y < \hat{q}_\tau\}\big)(y - \hat{q}_\tau),
\end{equation}
where $y$ denotes the observed realization and $\hat{q}_\tau$ the predicted $\tau$-quantile. Lower values indicate higher accuracy, as the score jointly reflects calibration and sharpness of the forecast distribution. 

Since forecasts are generated on a dense grid of 201 quantiles, we aggregate the pinball scores across all quantiles. The resulting sum is treated as an approximation of the \textit{Continuous Ranked Probability Score} (CRPS), which provides a single summary measure of the overall accuracy of the predictive distribution. The reported CRPS values are computed over the period from 31.12.2015 to 31.12.2023. 

In addition, we apply the Giacomini and White \cite{gia:whi:06} conditional predictive ability (CPA) test to assess whether the differences in CRPS between the quantile regression (QR) and historical simulation (HS) methods are statistically significant.
}

\subsubsection{CRPS results}

{Table \ref{tab:CRPS} reports the aggregated CRPS values for probabilistic forecasts of load, solar, wind, RES, and ResLoad. Across all variables, the quantile regression (QR) method achieves the lowest CRPS, with the differences relative to historical simulation (HS) and conformal prediction (CP) being statistically significant at the 1\% level according to the CPA \cite{gia:whi:06} test. This confirms that QR provides the most accurate and well-calibrated probabilistic forecasts of the fundamental variables. Moreover, QR is the only method that is adaptive -- its prediction intervals widen when the expected level of the variable increases, reflecting the higher uncertainty associated with larger load or generation forecasts. In contrast, the HS and CP methods produce nearly constant interval widths from day to day, differing mainly in location rather than shape, which likely limits their usefulness as dynamic inputs in the subsequent electricity price forecasting models. Consequently, QR-based probabilistic forecasts are not only more precise but also more suitable as inputs to electricity price forecasting models.}

\begin{table}[]
\caption{{Continuous Ranked Probability Score (CRPS) for probabilistic forecasts of load, solar, wind, RES, and residual load (ResLoad) obtained with historical simulation (HS), conformal prediction (CP) and quantile regression (QR). Asterisks indicate values significantly different from QR at the 1\% level according to the CPA test.}}
\label{tab:CRPS}
\begin{tabular}{r|ccccc}
 & Load & Solar & Wind & RES & ResLoad \\ 
 \hline
HS & 0.6275* & 0.1329* & 0.4286* & 0.4744* & 0.7954* \\
CP & 0.7730* & 0.1331* & 0.4296* & 0.4755* & 0.9071* \\
QR & \textbf{0.6103} & \textbf{0.1301} & \textbf{0.4160} & \textbf{0.4676} & \textbf{0.7869}
\end{tabular}
\end{table}

\subsection{Accuracy of electricity price forecasts}

\subsubsection{Evaluation measures}

Linear measures, such as mean absolute error (MAE), provide a simple, easy-to-understand assessment of average forecasting error. However, {the MAE measure is} intended to evaluate the accuracy of median forecasts, not mean predictions, {see e.g. \cite{gneiting2011making} for more details. However, the mean (or the expected value) is usually regarded as the key characteristic of interest of market, including electricity markets. }Therefore, in this study, {as the core evaluation metric we selected a} root mean square error (RMSE), {backed up with MAE based evaluation metrics.} The RMSE and MAE for a single day is defined by the following equation:

\begin{equation}
{
\text{RMSE}_d = \sqrt{\frac{1}{24}\sum_{h=1}^{24} \hat{\varepsilon}_{d,h}^2}
\quad\text{and}\quad 
\text{MAE}_d = \frac{1}{24}\sum_{h=1}^{24} |\hat{\varepsilon}_{d,h}|
.
}
\label{eqn:RMSEd}
\end{equation}

where $\hat{\varepsilon}_{d,h} = {p_{d,h} -\hat{p}_{d,h}}$ is the forecasting error for day $d$ and hour $h$. The aggregated {measures for the whole out-of-sample period of $D$ days are} defined as follows: 
\begin{equation}
{
\text{RMSE} = \sqrt{\frac{1}{D}\sum_{d=1}^{D} \text{RMSE}_d^2}
\quad\text{and}\quad 
\text{MAE} = {\frac{1}{D}\sum_{d=1}^{D} \text{MAE}_d}
}
\end{equation}

In addition, we perform the conditional predictive ability \citep[CPA;][]{gia:whi:06} test to formally evaluate the performance of the considered models, by pairwise evaluation. For two selected models A and B, we test the null hypothesis $H_0:\boldsymbol{\phi}=0$ in the following regression:
\begin{equation}\label{eqn:GW}
\Delta_d^{\text{A}, \text{B}}= \phi_0 + \phi_1 \Delta_{d-1}^{\text{A}, \text{B}} + \epsilon_d^{\text{A}, \text{B}}, 
\end{equation}
where $\Delta_{d}^{\text{A}, \text{B}} = \text{RMSE}_{d}^{\text{A}} - \text{RMSE}_{d}^{\text{B}}$, is the loss differential series and $\text{RMSE}_{d}^i$ is the prediction errors of model $i$ for day $d$.

\subsubsection{Comparison of different sets of fundamental variable}

Since several factors affect the results, we compare the accuracy of the predictions obtained for different subsets of models. The global comparison of all models is available in the appendix in terms of RMSE in Tables \ref{tab:app} and \ref{tab:app2} and in terms of MAE in Tables \ref{tab:app3} and \ref{tab:app4}. Note that the latter one are qualitatively similar to the one obtained with RMSE. 

\begin{figure*}[htb!]
 \includegraphics[width = 0.99\linewidth]{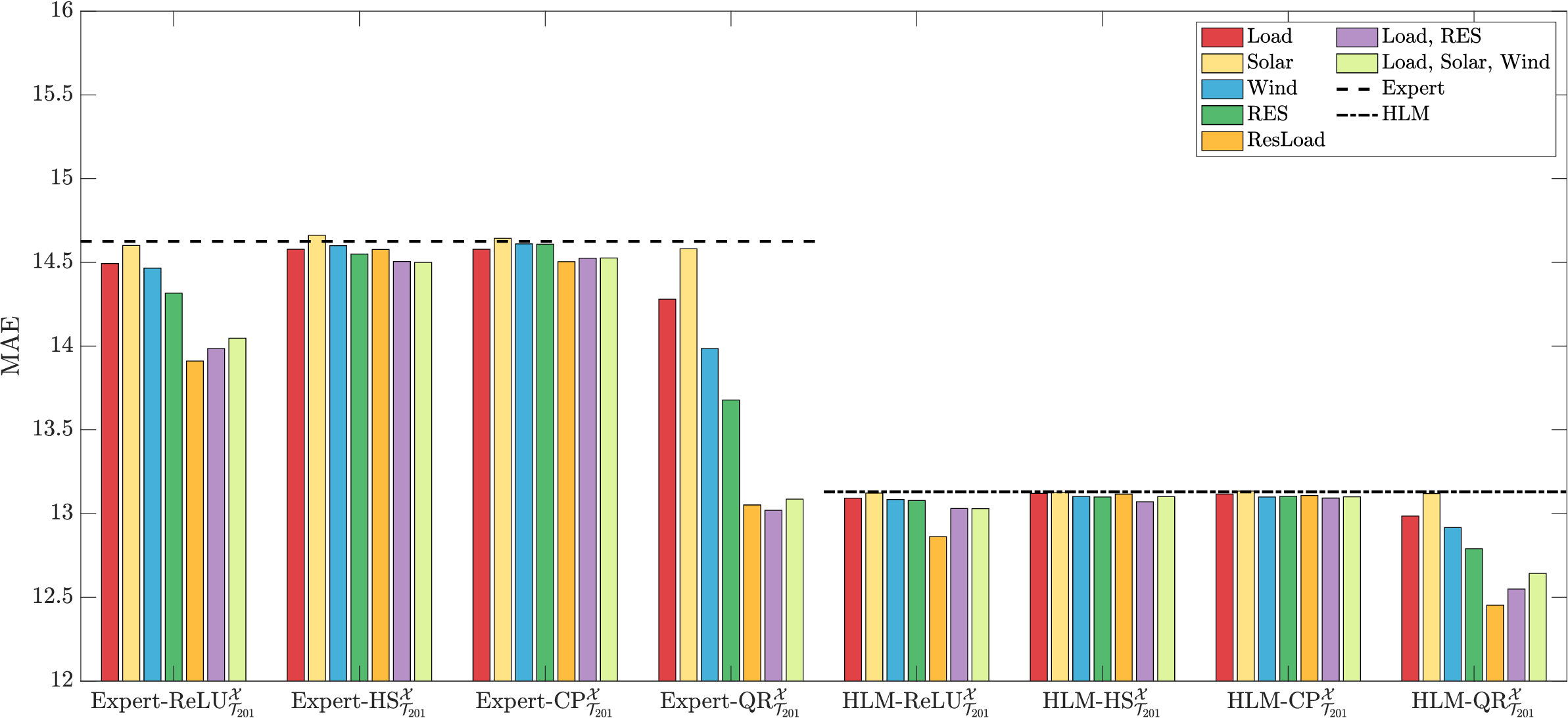}
 \caption{{Mean absolute error (MAE) of electricity price forecasts obtained from models using different sets of probabilistic input variables: forecasts of load, solar, wind, RES, and residual load (ResLoad). The figure illustrates how the use of probabilistic inputs affects the accuracy of point price forecasts for models including all 201 quantiles ($\mathcal{T}_{201}$)}}
 \label{fig:MAEmodel}
\end{figure*}

In {Figures~\ref{fig:RMSEmodel} and \ref{fig:MAEmodel}, we report the RMSE and MAE, respectively,} for all proposed models that include the 201-quantile prediction of fundamental variables. Then, we compare these models to benchmarks that do not include any probabilistic inputs. We examine the accuracy of electricity price predictions across various base model structures and techniques for obtaining probabilistic predictions of fundamental variables. Finally, we compare predictions using different sets of these variables as probabilistic inputs. The results presented in {Figures~\ref{fig:RMSEmodel} and \ref{fig:MAEmodel}} allow us to draw some important conclusions:

 \begin{figure*}[htb!]
 \includegraphics[width = 0.99\linewidth]{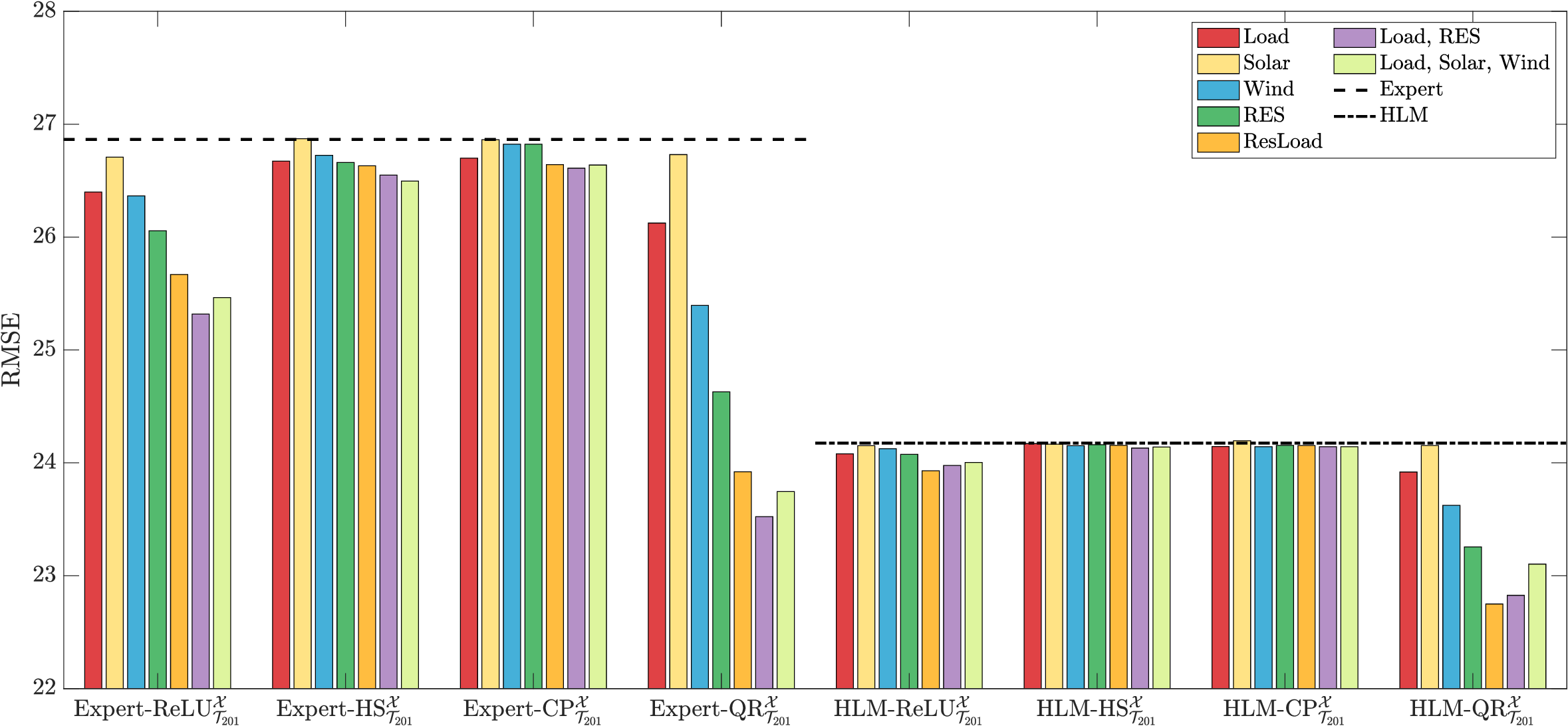}
 \caption{{Root mean square error (RMSE) of electricity price forecasts obtained from models using different sets of probabilistic input variables: forecasts of load, solar, wind, RES, and residual load (ResLoad). The figure illustrates how the use of probabilistic inputs affects the accuracy of point price forecasts for models including all 201 quantiles ($\mathcal{T}_{201}$)}}
 \label{fig:RMSEmodel}
\end{figure*}

\begin{itemize}
 \item The key insight is that models incorporating probabilistic forecasts of load and renewable generation consistently outperform those relying only on deterministic inputs. This demonstrates that RES/load uncertainty is not noise, but essential information for predicting prices in RES-integrated systems. This is true for both base models with higher improvement for the Expert model (up to 13.3\% in terms of RMSE and 11.6\% in terms of MAE; see Tables \ref{tab:app} and \ref{tab:app3} in Appendix).
 \item As expected, the parameter-rich (HLM) based models outperformed their Expert model based counterparts. However, some Expert models with probabilistic inputs (e.g., Expert-QR$^{\{\text{Load, RES}\}}_{\mathcal{T}_{201}}$) outperform HLM-based benchmark in terms of accuracy.
 \item Although any technique for obtaining probabilistic forecasts of the fundamental variables provides accuracy gains, by far the most accurate forecasts are obtained with probabilistic inputs calculated with QR. This indicates that the quality of the probabilistic forecast has a significant impact on the accuracy of the final electricity price forecasts.
 \item For all considered cases, the best performing models include probabilistic information on all three considered fundamental variables ($\mathcal{X} =$ \{ResLoad\}, $\mathcal{X} =$ \{Load, RES\} or $\mathcal{X} =$ \{Load, Solar, Wind\}). In particular, the best performing model is HML-QR$_{\mathcal{T}_{201}}^\text{ResLoad}$, which reduces the RMSE by 16.6\% and 6.1\% {(and MAE by 16.1\% and 5.3\%)} compared to the Expert and HLM benchmarks, respectively.
 \item {The results are consistent for both error measures highlighting the robustness of proposed approach for linear and quadratic measures}
\end{itemize}

\begin{figure}[hbt!]
 \includegraphics[width = 0.60\linewidth]{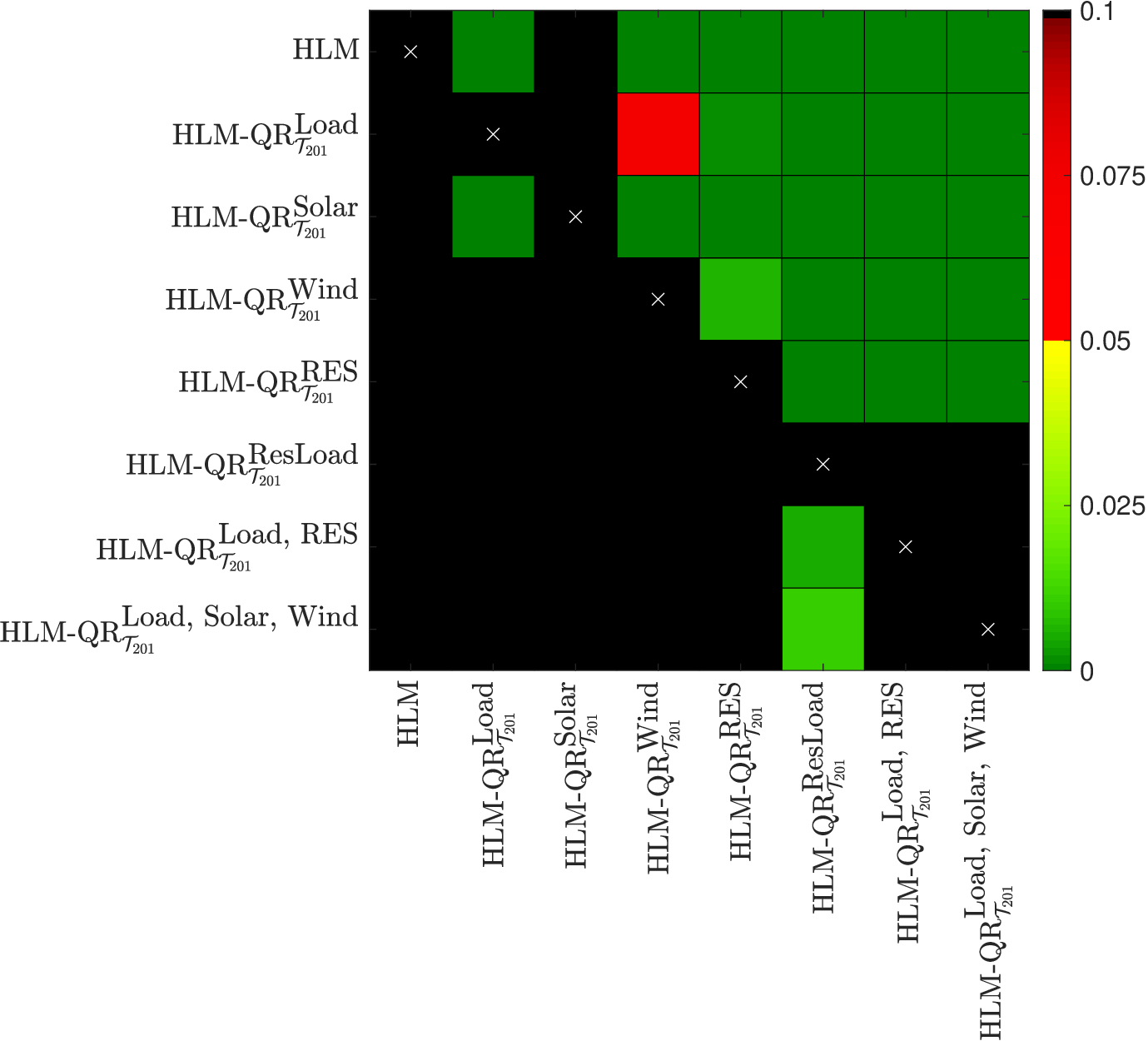}
 \caption{Results of the CPA test of \cite{gia:whi:06} for the RMSE.
 Heat maps are used to illustrate the range of $p$-values -- the smaller they are ($\rightarrow$ dark green), the more significant the difference between the two forecasts (the model on the X-axis outperforms the model on the Y-axis).}
 \label{fig:GWmodel}
\end{figure}

To formally confirm the significance of the difference between using different sets of fundamental variables included in the model in the form of probabilistic inputs in Figure~\ref{fig:GWmodel}, we present the results of the CPA test for the HLM benchmark and all HLM-QR$_{\mathcal{T}_{201}}^*$ models.

As can be seen in Figure~\ref{fig:GWmodel}, the HLM-QR$_{\mathcal{T}_{201}}^\text{ResLoad}$ models significantly outperform all other considered models. Overall, the three best models that significantly outperform other competitors include information about all three load, solar and wind (aggregated or not). On the other hand, the benchmark model without probabilistic inputs never outperforms any other model, while it performs significantly worse in 6 out of 7 cases, highlighting that including probabilistic inputs is beneficial in terms of forecast accuracy.

\subsubsection{Comparison of different number of quantiles}

\begin{figure*}[htb!]
\includegraphics[width = 0.99\linewidth]{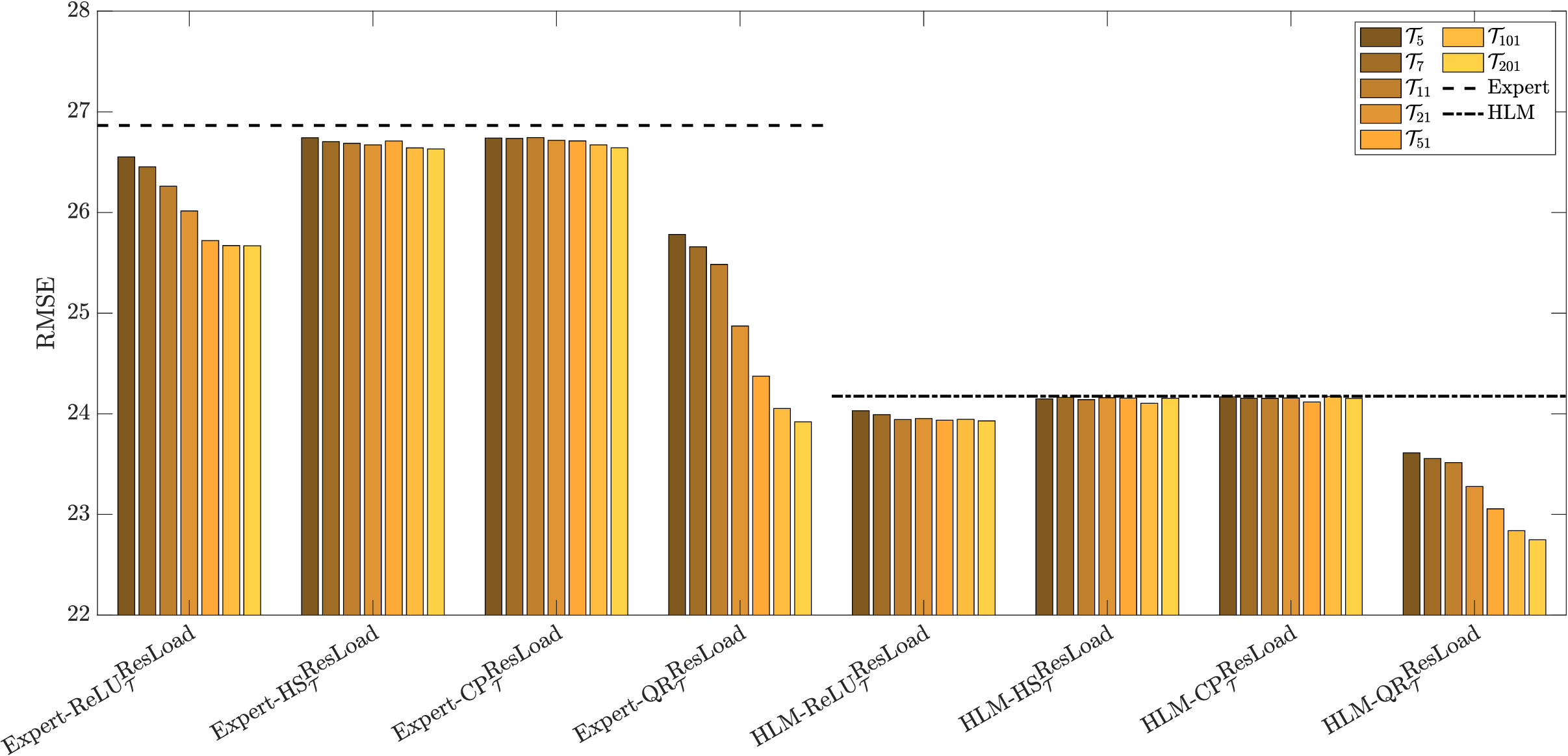}
 \caption{RMSE in EUR/MWh for the whole out-of-sample period for models with ResLoad probabilistic inputs}
 \label{fig:RMSEquantiles}
\end{figure*}

In Figure~\ref{fig:RMSEquantiles} we present a similar comparison as in Figure~\ref{fig:RMSEmodel}. However, this time we consider solely models including ResLoad and compare the forecast accuracy depending on the number of considered probability levels. Note that even though we only present the results for models consisting of ResLoad, the performance of other sets of variables are very similar (see Table \ref{tab:app}-\ref{tab:app4} in Appendix). 

Based on the results presented in Figure~\ref{fig:RMSEquantiles}, it can be seen that in general, the denser the grid of probability levels, the more accurate the predictions can be obtained. In particular, within the class of expert models, the best performing model is Expert-QR$_{\mathcal{T}_{201}}^\text{ResLoad}$, while its HLM-based equivalent is the best performing model within the class of parameter-rich models.

Furthermore, the results presented in Fig. \ref{fig:RMSEquantiles} confirm the findings observed in Fig. \ref{fig:RMSEmodel} and indicate that the models with probabilistic inputs generated with quantile regression are by far more accurate compared to those obtained with historical simulation or ReLU. 

\subsubsection{{Robustness check}}

{To verify that the results are not specific to a single evaluation metric, following the recommendation of \cite{lag:mar:des:wer:21} we complement the main analysis with relative accuracy measures: the relative mean absolute error (rMAE) and the relative root mean square error (rRMSE). These metrics normalize performance with respect to a naive benchmark, allowing for direct comparison across models and variables. Following \cite{lag:mar:des:wer:21} we define naive benchmark as $\hat{p}_{d,h} = p_{d-7,h}$, i.e., the observed price from the same hour one week earlier. The relative measures rRMSE and rMAE are defined as
\begin{equation}
\text{rRMSE} = \frac{\text{RMSE}_{\text{model}}}{\text{RMSE}_{\text{naive}}}
\quad\text{and}\quad 
\text{rMAE} = \frac{\text{MAE}_{\text{model}}}{\text{MAE}_{\text{naive}}} .
\end{equation}
}

{Table \ref{tab:relatives} presents the results of the robustness check using both absolute (RMSE, MAE) and relative (rRMSE, rMAE) accuracy measures, computed with respect to the weekly persistence benchmark. The naive model serves as a reference with relative scores equal to one. Across both the Expert and HLM classes, incorporating probabilistic inputs improves forecast accuracy, with the QR-based models consistently achieving the lowest RMSE and MAE as well as the smallest relative errors. The improvements are particularly visible in the HLM class, where the inclusion of QR-derived probabilistic forecasts reduces rRMSE to 0.41 and rMAE to 0.40. These results confirm that the gains obtained from probabilistic residual load forecasts are robust across different accuracy metrics and model architectures.}

\renewcommand{\arraystretch}{1.4}

\begin{table}[htb]
\caption{{Root mean square error (RMSE), mean absolute error (MAE), and their relative versions (rRMSE, rMAE) for selected best performing model specifications. Relative metrics are computed with respect to the weekly persistence naive benchmark $\hat{p}_{d,h} = p_{d-7,h}$.}}
\label{tab:relatives}
\begin{tabular}{r|cccc}
& RMSE & rRMSE & MAE & rMAE \\
\hline
{Naive}                                              & {56.66} & {1.00} & {30.66} & {1.00} \\
\hline
{Expert}                                             & {26.86} & {0.48} & {14.63} & {0.47} \\
{Expert-ReLU$_{\mathcal{T}_{201}}^{\mbox{ResLoad}}$} & {25.67} & {0.45} & {13.91} & {0.45} \\
{Expert-HS$_{\mathcal{T}_{201}}^{\mbox{ResLoad}}$}   & {26.63} & {0.48} & {14.58} & {0.47} \\
{Expert-CP$_{\mathcal{T}_{201}}^{\mbox{ResLoad}}$}   & {26.64} & {0.47} & {14.50} & {0.47} \\
{Expert-QR$_{\mathcal{T}_{201}}^{\mbox{ResLoad}}$}   & {23.92} & {0.43} & {13.05} & {0.42} \\
\hline
{HLM}                                                & {24.17} & {0.43} & {13.13} & {0.43} \\
{HLM-ReLU$_{\mathcal{T}_{201}}^{\mbox{ResLoad}}$}    & {23.93} & {0.42} & {12.86} & {0.42} \\
{HLM-HS$_{\mathcal{T}_{201}}^{\mbox{ResLoad}}$}      & {24.15} & {0.43} & {13.12} & {0.43} \\
{HLM-CP$_{\mathcal{T}_{201}}^{\mbox{ResLoad}}$}      & {24.15} & {0.43} & {13.11} & {0.43} \\
{HLM-QR$_{\mathcal{T}_{201}}^{\mbox{ResLoad}}$}      & {22.75} & {0.41} & {12.45} & {0.40}
\end{tabular}
\end{table}

\subsubsection{Performance improvement analysis}

\begin{figure}[htb!]
 \includegraphics[width = 0.8\linewidth]{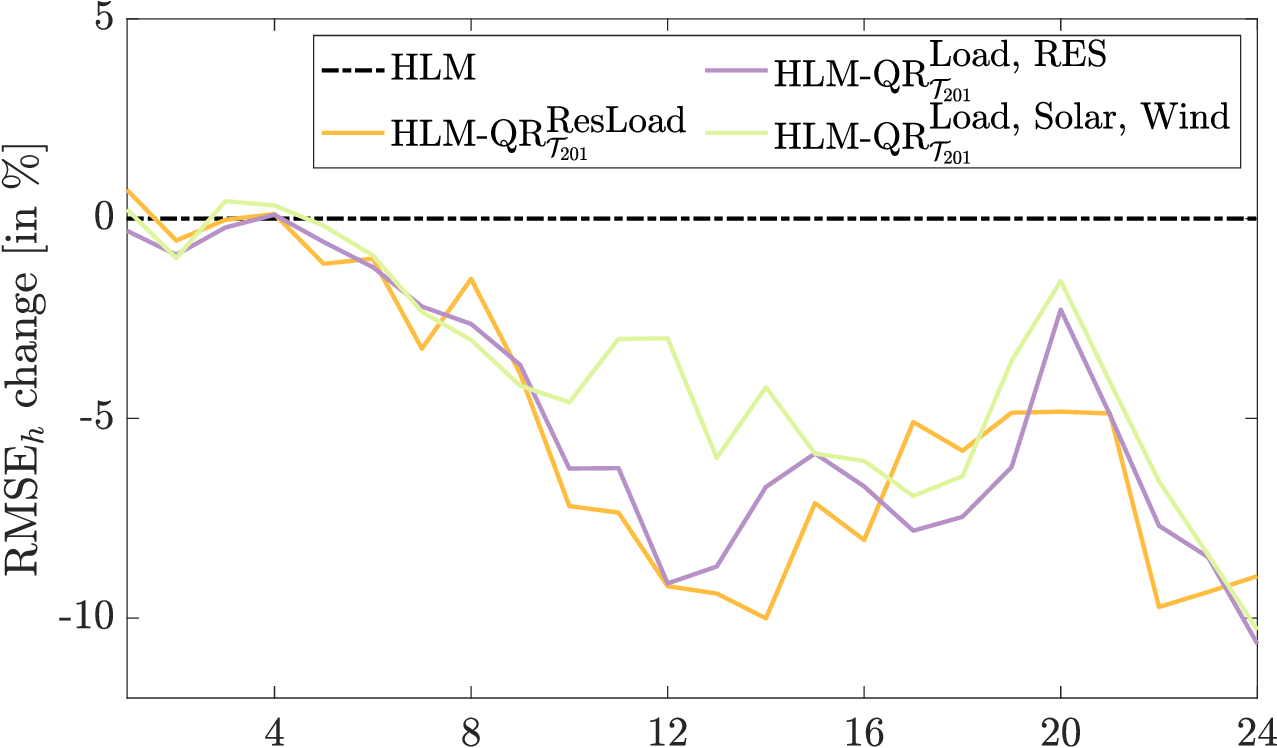}
 \caption{{Hourly root mean square error (RMSE$_h$) of day-ahead electricity price forecasts. Each value represents the RMSE$_h$ calculated separately for a given delivery hour (1–24) over the entire out-of-sample period. Percentage improvements over the HLM benchmark are presented for selected best performing models.}}
 \label{fig:HourlyRMSE}
\end{figure}

To analyze the performance of our newly proposed approach, in Figure~\ref{fig:HourlyRMSE} we present the {hour-specific RMSE percentage improvement. This hourly decomposition is standard in electricity price forecasting and allows us to capture intra-day differences in predictability across different hours of a day.} 
In Figure~\ref{fig:HourlyRMSE} we compare the performance of three top performing models (HLM-QR$^\text{ResLoad}_{\mathcal{T}_{201}}$, HLM-QR$^\text{Load, RES}_{\mathcal{T}_{201}}$, HLM-QR$^\text{Load, Solar, Wind}_{\mathcal{T}_{201}}$) against the HLM benchmark. The percentage improvement for hour $h$ is defined as 
\begin{equation}
 \text{\%chng}_h = \log \left( \text{RMSE}_h^{(i)} / \text{RMSE}_h^{\text{HLM}} \right)\times 100\%,
\end{equation}
where $\text{RMSE}_h^{(i)}=\sqrt{\frac{1}{D}\sum_{d=1}^{D} \hat{\varepsilon}_{d,h}^2}$ is the root mean square error {computed for a specific hour $h$} and obtained with model $i$.

As can be seen in Figure~\ref{fig:HourlyRMSE}, all three considered models outperform the benchmark for the majority of hours. Although the results of the models with different sets of probabilistic inputs are different, the overall trend remains the same. The improvement is smaller for the early morning hours (1-6) but increases to roughly 10\% for the midday and late evening hours (22-24). This shows that modeling load and RES uncertainty is most beneficial at times when the system is under stress, underscoring its importance for market stability and renewable integration.

\subsubsection{Selection and impact of probabilistic inputs}
\label{sec:impact}

To answer the question of what is the most important factor in our approach that leads to a significant improvement in accuracy, we can further analyze how often each variable was selected in the final model and how given variables affect the prediction.

\begin{figure}[htb!]
 \includegraphics[width = 0.75\linewidth]{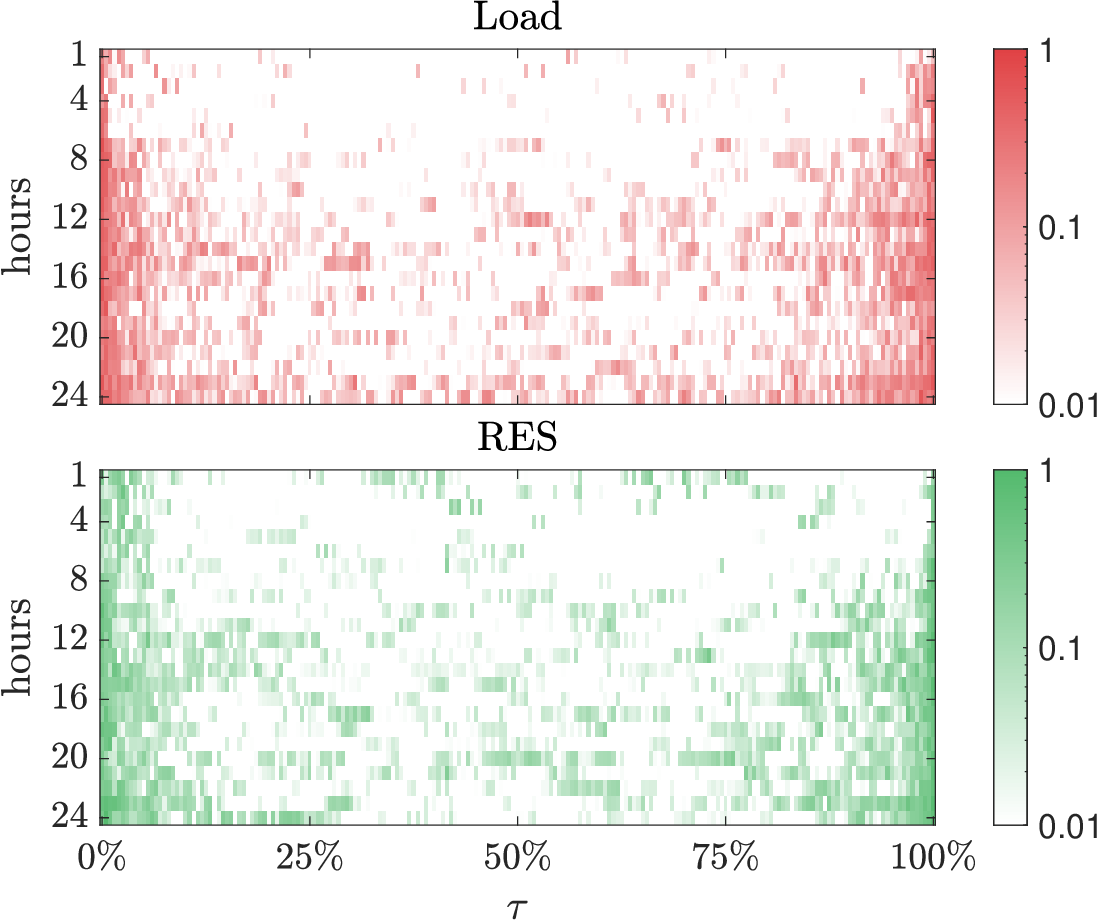}
 \caption{The plots show how often (in percentage) given quantile predictions for Load and RES are included in the final HLM-QR$_{\mathcal{T}_{201}}^{Load, RES}$ model (corresponding $\beta$ has non-zero value). We report the percentages separately for each probability level $\tau = 0.005, 0.01, \ldots, 0.995$ (Y-axis) and for each hour of a day (X-axis).}
 \label{fig:Frequency}
\end{figure}

In Fig. \ref{fig:Frequency} we show how often a given quantile prediction of load and RES was selected in the final model (corresponding $\beta$ value was different from 0). The results are presented for the HLM-QR$_{\mathcal{T}_{201}}^{\text{Load}, \text{RES}}$ model. The darker the color in Fig. \ref{fig:Frequency}, the more often the quantile prediction was selected as important by the LASSO regularization at the given probability level. It can be seen that by far the darkest places are on the left and right side of both plots, indicating that the extreme quantiles are selected much more often than the center of the distribution. This indicates that rare but system-critical scenarios — such as very low wind or sudden demand spikes — are disproportionately important for price formation. It confirms that electricity prices in renewable-rich systems are driven less by average conditions and more by the tails of the load and RES distributions. Another interesting observation is that the probabilistic inputs are generally selected more frequently in the afternoon and evening hours and very rarely in the morning hours.

\begin{figure*}[htb!]
 \includegraphics[width = 0.99\linewidth]{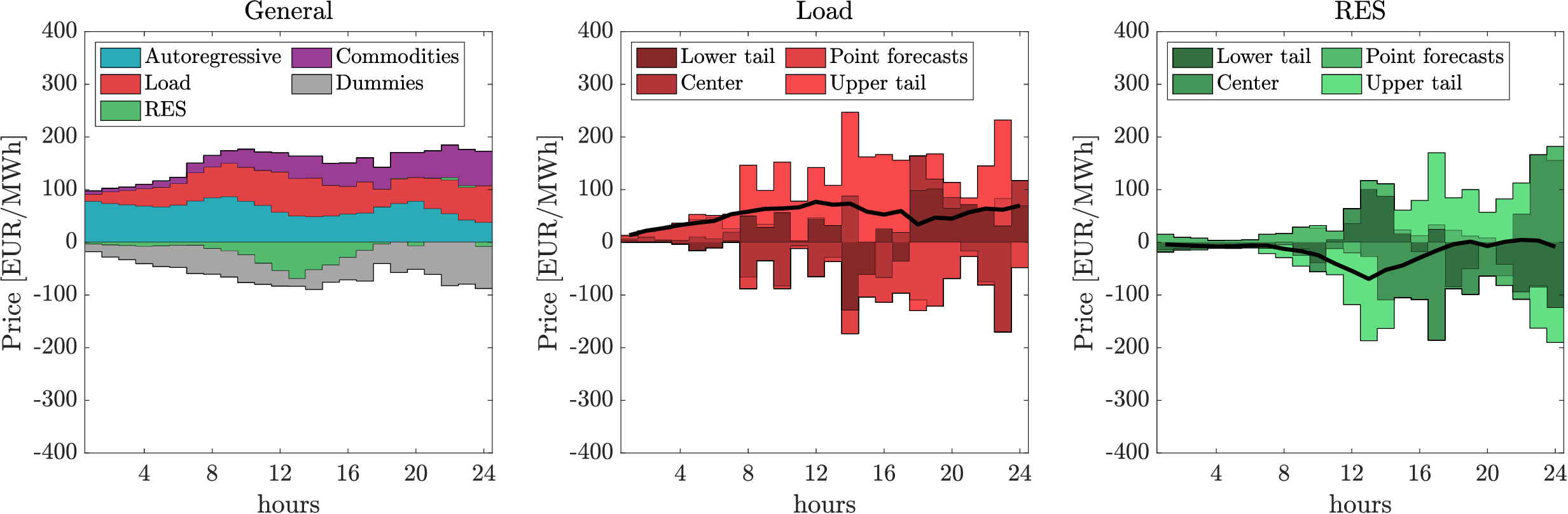}
 \caption{Average impact of variable groups on electricity price forecasts using the HLM-QR${\mathcal{T}{201}}^{Load, RES}$ model. General variable groups (left panel) include autoregressive, load, RES, commodities, and intercept terms. The middle and right panels show the influence of different parts of the load and RES distributions. The black lines in the middle and right plots represent the general influence of Load and RES, respectively}
 \label{fig:Impact}
\end{figure*}

To analyze the impact of the probabilistic inputs on the forecast accuracy in Figure~\ref{fig:Impact} we report the average impact of a given group of variables on the price forecast obtained with the HLM-QR$_{\mathcal{T}_{201}}^{\text{Load}, \text{RES}}$ model. We quantify the impact of variable groups by measuring the average contribution of each group to the final price prediction. This provides insight into which types of inputs—autoregressive, load, RES, commodities—play the largest roles at different hours.

First, in the left panel of Figure~\ref{fig:Impact} we plot the impact of 5 general groups of variables:
\begin{itemize}[itemsep=0pt]
 \item Autoregressive part consisting of 50 variables related to historical electricity prices ($\beta_1$ to $\beta_{50}$ in Eq.\eqref{eq:HLM}),
 \item Load, which includes 48 variables related to the day-ahead load point forecasts ($\beta_{51}$ to $\beta_{98}$ in equation \eqref{eq:HLM}) plus 199 quantile load forecasts. 
 \item RES which includes 96 variables related to day-ahead point forecasts of solar and wind generation ($\beta_{99}$ to $\beta_{194}$ in Eq. \eqref{eq:HLM}) plus 199 quantile forecasts of RES
 \item Commodities, including 4 variables ($\beta_{195}$ to $\beta_{198}$ in equation \eqref{eq:HLM})
 \item Intercept, including 7 dummy variables ($\beta_{199}$ to $\beta_{205}$ in eq. \eqref{eq:HLM}) and intercept.
\end{itemize}
The middle and right panels in Figure~\ref{fig:Impact} show the impact of different parts of the distribution of predicted load and RES. The lower and upper tails are each represented by 21 quantile predictions (ranging from $\gamma$ to 10\% and from 90\% to $1-\gamma$, respectively), while the middle of the distribution consists of 159 variables associated with quantile predictions with probability levels ranging from 10.5\% to 74.5\%.

Based on Figure~ \ref{fig:Impact} it can be seen that the proposed model exhibits stronger autoregressive behavior at the beginning of the day and weaker autoregressive behavior toward the end of the day, allowing for a greater influence of fundamental factors such as fuel prices. This is consistent with the current literature \citep{ghelasi2025day}. However, the influence of RES seems to be significantly reduced, as it is mainly active in the middle of the day (probably related to solar generation). 

For the middle and right plots in Figure~\ref{fig:Impact}, it is evident that the impact is minimal during the early morning hours. This is likely due to the dominance of autoregressive information, with fundamental factors providing limited additional insight. Throughout the distribution -- left tail, right tail, and center -- each component is shown to be relevant, but their relative importance changes over time in a non-trivial manner.

\section{Conclusions}
\label{sec:conclusion}
This study examined whether electricity price forecasts can be improved by integrating probabilistic forecasts of load, wind, and solar generation into established day-ahead price models. Using German EPEX data for 2015–2023, we demonstrated that both parsimonious expert models and high-dimensional LASSO specifications consistently benefit from probabilistic inputs. 

Our analysis shows that the integration of probabilistic inputs significantly improves forecast accuracy. Among the parameter-rich HLM-based models and their expert counterparts, the HLM-QR$_{\mathcal{T}_{201}}^\text{ResLoad}$ model performs best, reducing the RMSE by 16.5 \% and 6.1\% over the expert and HLM benchmarks, respectively. 

In addition, the quality of the probabilistic predictions of the fundamental variables strongly affects the performance of the models, with quantile regression (QR) being the most successful. moreover the results indicate that the LASSO procedure frequently selected extreme quantiles of load and renewable forecasts, suggesting that electricity prices are affected by rare yet critical events, such as sudden demand spikes, low wind, and renewable surpluses. These results highlight the importance of modeling the full distribution of fundamentals instead of relying solely on expected values. 

Detailed analysis of quantile inclusion revealed that denser grids of quantiles consistently improved forecast accuracy, highlighting the importance of research in probabilistic forecasting of load and RES generation. Particularly, in wind and solar power forecasting, probabilistic forecasts are hardly used in practice for decision making. Even if probabilistic forecasts are utilized, this is usually only for very limited quantiles, e.g. 10\%, 50\% and 90\% while a denser information set would be significantly better. These findings suggest that good quality of probabilistic forecasts of load and RES generation can be useful for a broad range of stakeholders. System operators can better anticipate imbalances and optimize reserve allocation; traders and utilities can improve bidding strategies and hedge against renewable uncertainty; and policymakers may view probabilistic forecasting as a tool to reduce integration costs and facilitate the energy transition.

The results indicate that the performance improvements were not uniform across all hours: while the gains were relatively small in the early hours of the day, they were very substantial in the late hours of the day. This variability is consistent with the observation that probabilistic inputs are selected more frequently in the afternoon and evening hours, particularly extreme quantiles, which are more relevant at these times.

This study focuses on linear models (Expert and HLM) to isolate the effect of probabilistic input variables. However, our findings can serve as a benchmark for future work on non-linear architectures (e.g., deep learning or gradient boosting) incorporating similar probabilistic structures. We anticipate that nonlinear models could benefit even further from rich probabilistic features, but that lies beyond the scope of this analysis.

While the main focus of this paper is to improve point forecasts of electricity prices using probabilistic inputs, a natural extension of this approach would involve generating full probabilistic forecasts of prices, particularly using quantile regression on the price itself. This topic is left for future research.

\section*{Acknowledgments}
The study was partially supported by the National Science Center (NCN, Poland) through grant no.\ 2023/49/N/HS4/02741 (to BU). This research was partially funded in the course of TRR 391 Spatio-temporal Statistics for the Transition of Energy and Transport (520388526) by the Deutsche Forschungsgemeinschaft (DFG,
German Research Foundation) (to FZ). 

BU is supported by the Foundation for Polish Science (FNP)

A part of this work was developed during BU’s research stay at the University of Duisburg-Essen, Germany, in August and September 2025.

\section*{Data availability}
The raw data required to reproduce the above findings are available to download from \url{transparency.entsoe.eu} and \url{investing.com}. 

\bibliographystyle{elsarticle-num}
\bibliography{EPF}

\appendix
\renewcommand{\arraystretch}{1.2}

\setcounter{table}{0}
\section{Root mean square error for all considered models}

\begin{table*}[htb!]
\caption{{Root mean square error and percentage improvement over the benchmark (\%chng) of Expert-based forecasting models. The reported percentage change refers to the improvement over the model without any probabilistic inputs. The green color indicates the best performing models.}}
\label{tab:app}
\scalebox{0.7}{
\begin{tabular}{rrr|cr|cr|cr|cr|cr|cr|cr}
 & \multicolumn{1}{l}{--} & -- & \multicolumn{14}{c}{26.86} \\
 \multicolumn{1}{l}{} & \multicolumn{1}{l}{} && \multicolumn{2}{c|}{$\mathcal{T}_5$} & \multicolumn{2}{c|}{$\mathcal{T}_7$} & \multicolumn{2}{c|}{$\mathcal{T}_{11}$} & \multicolumn{2}{c|}{$\mathcal{T}_{21}$} & \multicolumn{2}{c|}{$\mathcal{T}_{51}$} & \multicolumn{2}{c|}{$\mathcal{T}_{101}$} & \multicolumn{2}{c}{$\mathcal{T}_{201}$} \\
 &&& \multicolumn{1}{c}{\rotatebox{90}{RMSE}} & \multicolumn{1}{c|}{\rotatebox{90}{\%chng}} & \multicolumn{1}{c}{\rotatebox{90}{RMSE}} & \multicolumn{1}{c|}{\rotatebox{90}{\%chng}} & \multicolumn{1}{c}{\rotatebox{90}{RMSE}} & \multicolumn{1}{c|}{\rotatebox{90}{\%chng}} & \multicolumn{1}{c}{\rotatebox{90}{RMSE}} & \multicolumn{1}{c|}{\rotatebox{90}{\%chng}} & \multicolumn{1}{c}{\rotatebox{90}{RMSE}} & \multicolumn{1}{c|}{\rotatebox{90}{\%chng}} & \multicolumn{1}{c}{\rotatebox{90}{RMSE}} & \multicolumn{1}{c|}{\rotatebox{90}{\%chng}} & \multicolumn{1}{c}{\rotatebox{90}{RMSE}} & \multicolumn{1}{c}{\rotatebox{90}{\%chng}} \\
 \cline{2-17}
 \noalign{\smallskip}
&& Load& 26.78 & \cellcolor[HTML]{F6F9F9}-0.3 & 26.84 & \cellcolor[HTML]{F8FAFC}-0.1 & 26.80 & \cellcolor[HTML]{F6F9FA}-0.3 & 26.69 & \cellcolor[HTML]{F2F8F6}-0.7& 26.52 & \cellcolor[HTML]{EBF5F0}-1.3& 26.43 & \cellcolor[HTML]{E7F3ED}-1.6& 26.40 & \cellcolor[HTML]{E5F3EB}-1.8\\
&& Solar & 26.80 & \cellcolor[HTML]{F7FAFA}-0.2 & 26.81 & \cellcolor[HTML]{F7FAFB}-0.2 & 26.81 & \cellcolor[HTML]{F7FAFA}-0.2 & 26.74 & \cellcolor[HTML]{F4F8F8}-0.5& 26.71 & \cellcolor[HTML]{F3F8F7}-0.6& 26.68 & \cellcolor[HTML]{F1F7F6}-0.7& 26.71 & \cellcolor[HTML]{F3F8F7}-0.6\\
&& Wind& 26.86 & \cellcolor[HTML]{F9FAFC}0.0& 26.90 & \cellcolor[HTML]{FBFBFE}0.1& 26.83 & \cellcolor[HTML]{F8FAFB}-0.1 & 26.72 & \cellcolor[HTML]{F3F8F7}-0.6& 26.58 & \cellcolor[HTML]{EDF6F2}-1.1& 26.43 & \cellcolor[HTML]{E7F3ED}-1.6& 26.37 & \cellcolor[HTML]{E4F2EA}-1.9\\
&& RES& 26.73 & \cellcolor[HTML]{F3F8F7}-0.5 & 26.63 & \cellcolor[HTML]{EFF7F4}-0.9 & 26.59 & \cellcolor[HTML]{EEF6F3}-1.0 & 26.41 & \cellcolor[HTML]{E6F3EC}-1.7& 26.20 & \cellcolor[HTML]{DDEFE4}-2.5& 26.10 & \cellcolor[HTML]{D8EDE0}-2.9& 26.06 & \cellcolor[HTML]{D7EDDF}-3.1\\
&& ResLoad& 26.55 & \cellcolor[HTML]{ECF5F1}-1.2 & 26.45 & \cellcolor[HTML]{E8F3ED}-1.5 & 26.26 & \cellcolor[HTML]{DFF0E6}-2.3 & 26.02 & \cellcolor[HTML]{D5ECDD}-3.2& 25.72 & \cellcolor[HTML]{C8E7D2}-4.4& 25.67 & \cellcolor[HTML]{C6E6D0}-4.5& 25.67 & \cellcolor[HTML]{C6E6D0}-4.6\\
&& Load, RES& 26.62 & \cellcolor[HTML]{EFF6F3}-0.9 & 26.62 & \cellcolor[HTML]{EFF6F4}-0.9 & 26.54 & \cellcolor[HTML]{EBF5F1}-1.2 & 26.25 & \cellcolor[HTML]{DFF0E6}-2.3& 25.78 & \cellcolor[HTML]{CAE8D4}-4.1& 25.52 & \cellcolor[HTML]{BFE3CA}-5.1& 25.32 & \cellcolor[HTML]{B6DFC2}-5.9\\
& \multirow{-7}{*}{\rotatebox{90}{ReLU}} & Load, Solar, Wind & 26.73 & \cellcolor[HTML]{F3F8F8}-0.5 & 26.80 & \cellcolor[HTML]{F6F9FA}-0.2 & 26.69 & \cellcolor[HTML]{F2F8F6}-0.6 & 26.34 & \cellcolor[HTML]{E3F1E9}-2.0& 25.99 & \cellcolor[HTML]{D4EBDC}-3.3& 25.63 & \cellcolor[HTML]{C4E5CE}-4.7& 25.46 & \cellcolor[HTML]{BCE2C8}-5.4\\

\cline{2-17}
\noalign{\smallskip}
 
&& Load& 26.66 & \cellcolor[HTML]{F1F7F5}-0.7 & 26.67 & \cellcolor[HTML]{F1F7F5}-0.7 & 26.66 & \cellcolor[HTML]{F1F7F5}-0.8 & 26.63 & \cellcolor[HTML]{EFF6F4}-0.9& 26.59 & \cellcolor[HTML]{EDF6F2}-1.0& 26.61 & \cellcolor[HTML]{EFF6F3}-0.9& 26.67 & \cellcolor[HTML]{F1F7F5}-0.7\\
&& Solar & 26.89 & \cellcolor[HTML]{FAFBFD}0.1& 26.87 & \cellcolor[HTML]{FAFBFD}0.0& 26.89 & \cellcolor[HTML]{FAFBFD}0.1& 26.89 & \cellcolor[HTML]{FAFBFD}0.1& 26.89 & \cellcolor[HTML]{FAFBFE}0.1& 26.87 & \cellcolor[HTML]{FAFBFD}0.0& 26.87 & \cellcolor[HTML]{F9FBFD}0.0\\
&& Wind& 26.83 & \cellcolor[HTML]{F8FAFB}-0.1 & 26.83 & \cellcolor[HTML]{F8FAFB}-0.1 & 26.83 & \cellcolor[HTML]{F8FAFB}-0.1 & 26.83 & \cellcolor[HTML]{F8FAFB}-0.1& 26.76 & \cellcolor[HTML]{F5F9F9}-0.4& 26.74 & \cellcolor[HTML]{F4F8F8}-0.5& 26.72 & \cellcolor[HTML]{F3F8F7}-0.5\\
&& RES& 26.81 & \cellcolor[HTML]{F7FAFB}-0.2 & 26.80 & \cellcolor[HTML]{F7FAFA}-0.2 & 26.76 & \cellcolor[HTML]{F5F9F9}-0.4 & 26.75 & \cellcolor[HTML]{F4F9F8}-0.4& 26.73 & \cellcolor[HTML]{F4F8F8}-0.5& 26.69 & \cellcolor[HTML]{F2F8F6}-0.7& 26.66 & \cellcolor[HTML]{F1F7F5}-0.8\\
&& ResLoad& 26.74 & \cellcolor[HTML]{F4F8F8}-0.5 & 26.70 & \cellcolor[HTML]{F2F8F7}-0.6 & 26.69 & \cellcolor[HTML]{F2F8F6}-0.7 & 26.67 & \cellcolor[HTML]{F1F7F5}-0.7& 26.71 & \cellcolor[HTML]{F3F8F7}-0.6& 26.64 & \cellcolor[HTML]{F0F7F4}-0.8& 26.63 & \cellcolor[HTML]{EFF7F4}-0.9\\
&& Load, RES& 26.64 & \cellcolor[HTML]{EFF7F4}-0.9 & 26.63 & \cellcolor[HTML]{EFF6F4}-0.9 & 26.64 & \cellcolor[HTML]{F0F7F4}-0.9 & 26.60 & \cellcolor[HTML]{EEF6F3}-1.0& 26.57 & \cellcolor[HTML]{EDF6F2}-1.1& 26.55 & \cellcolor[HTML]{ECF5F1}-1.2& 26.55 & \cellcolor[HTML]{ECF5F1}-1.2\\
& \multirow{-7}{*}{\rotatebox{90}{HS}}& Load, Solar, Wind & 26.70 & \cellcolor[HTML]{F2F8F6}-0.6 & 26.69 & \cellcolor[HTML]{F2F8F6}-0.7 & 26.67 & \cellcolor[HTML]{F1F7F5}-0.7 & 26.66 & \cellcolor[HTML]{F0F7F5}-0.8& 26.55 & \cellcolor[HTML]{ECF5F1}-1.2& 26.55 & \cellcolor[HTML]{ECF5F1}-1.2& 26.50 & \cellcolor[HTML]{EAF4EF}-1.4\\
\cline{2-17}
\noalign{\smallskip}

 & & Load & 26.71 & \cellcolor[HTML]{F3F8F7}-0.6 & 26.7 & \cellcolor[HTML]{F4F8F8}-0.5 & 26.72 & \cellcolor[HTML]{F3F8F7}-0.5 & 26.7 & \cellcolor[HTML]{F2F8F7}-0.6 & 26.70 & \cellcolor[HTML]{F2F8F7}-0.6 & 26.7 & \cellcolor[HTML]{F1F7F5}-0.7 & 26.70 & \cellcolor[HTML]{F2F8F6}-0.6 \\
 & & Solar & 26.86 & \cellcolor[HTML]{F9FBFC}0.0 & 26.9 & \cellcolor[HTML]{F9FAFC}-0.1 & 26.86 & \cellcolor[HTML]{F9FAFC}0.0 & 26.9 & \cellcolor[HTML]{F9FAFC}0.0 & 26.84 & \cellcolor[HTML]{F8FAFC}-0.1 & 26.9 & \cellcolor[HTML]{F9FAFC}0.0 & 26.86 & \cellcolor[HTML]{F9FBFC}0.0 \\
 & & Wind & 26.92 & \cellcolor[HTML]{FBFBFE}0.2 & 26.9 & \cellcolor[HTML]{FCFCFF}0.2 & 26.92 & \cellcolor[HTML]{FBFBFE}0.2 & 26.9 & \cellcolor[HTML]{FBFBFE}0.1 & 26.91 & \cellcolor[HTML]{FBFBFE}0.2 & 26.9 & \cellcolor[HTML]{F9FBFD}0.0 & 26.82 & \cellcolor[HTML]{F7FAFB}-0.2 \\
 & & RES & 26.90 & \cellcolor[HTML]{FBFBFE}0.1 & 26.9 & \cellcolor[HTML]{FBFBFE}0.2 & 26.91 & \cellcolor[HTML]{FBFBFE}0.1 & 26.9 & \cellcolor[HTML]{FAFBFD}0.0 & 26.87 & \cellcolor[HTML]{F9FBFD}0.0 & 26.8 & \cellcolor[HTML]{F8FAFC}-0.1 & 26.82 & \cellcolor[HTML]{F7FAFB}-0.2 \\
 & & ResLoad & 26.74 & \cellcolor[HTML]{F4F8F8}-0.5 & 26.7 & \cellcolor[HTML]{F4F8F8}-0.5 & 26.74 & \cellcolor[HTML]{F4F9F8}-0.5 & 26.7 & \cellcolor[HTML]{F3F8F7}-0.6 & 26.71 & \cellcolor[HTML]{F3F8F7}-0.6 & 26.7 & \cellcolor[HTML]{F1F7F5}-0.7 & 26.64 & \cellcolor[HTML]{F0F7F4}-0.8 \\
 & & Load, RES & 26.74 & \cellcolor[HTML]{F4F8F8}-0.5 & 26.8 & \cellcolor[HTML]{F5F9F9}-0.4 & 26.74 & \cellcolor[HTML]{F4F8F8}-0.5 & 26.7 & \cellcolor[HTML]{F2F8F6}-0.6 & 26.67 & \cellcolor[HTML]{F1F7F5}-0.7 & 26.6 & \cellcolor[HTML]{EFF7F4}-0.9 & 26.61 & \cellcolor[HTML]{EEF6F3}-1.0 \\
 & \multirow{-7}{*}{\rotatebox{90}{CP}} & Load, Solar, Wind & 26.77 & \cellcolor[HTML]{F5F9F9}-0.4 & 26.8 & \cellcolor[HTML]{F5F9F9}-0.4 & 26.76 & \cellcolor[HTML]{F5F9F9}-0.4 & 26.7 & \cellcolor[HTML]{F4F8F8}-0.5 & 26.68 & \cellcolor[HTML]{F2F7F6}-0.7 & 26.7 & \cellcolor[HTML]{F1F7F6}-0.7 & 26.64 & \cellcolor[HTML]{F0F7F4}-0.8 \\

 \cline{2-17}
\noalign{\smallskip}

 && Load& 26.51 & \cellcolor[HTML]{EAF4EF}-1.3 & 26.44 & \cellcolor[HTML]{E7F3ED}-1.6 & 26.43 & \cellcolor[HTML]{E7F3EC}-1.6 & 26.25 & \cellcolor[HTML]{DFF0E6}-2.3& 26.16 & \cellcolor[HTML]{DBEEE3}-2.6& 26.11 & \cellcolor[HTML]{D9EDE1}-2.9& 26.12 & \cellcolor[HTML]{D9EEE1}-2.8\\
&& Solar & 26.83 & \cellcolor[HTML]{F8FAFB}-0.1 & 26.82 & \cellcolor[HTML]{F7FAFB}-0.2 & 26.78 & \cellcolor[HTML]{F6F9FA}-0.3 & 26.73 & \cellcolor[HTML]{F4F8F8}-0.5& 26.72 & \cellcolor[HTML]{F3F8F7}-0.5& 26.74 & \cellcolor[HTML]{F4F8F8}-0.5& 26.73 & \cellcolor[HTML]{F4F8F8}-0.5\\
&& Wind& 26.29 & \cellcolor[HTML]{E1F1E7}-2.2 & 26.22 & \cellcolor[HTML]{DDEFE5}-2.4 & 26.07 & \cellcolor[HTML]{D7EDDF}-3.0 & 25.93 & \cellcolor[HTML]{D1EADA}-3.6& 25.67 & \cellcolor[HTML]{C6E6D0}-4.5& 25.55 & \cellcolor[HTML]{C0E4CB}-5.0& 25.39 & \cellcolor[HTML]{B9E1C5}-5.6\\
&& RES& 25.75 & \cellcolor[HTML]{C9E7D3}-4.2 & 25.65 & \cellcolor[HTML]{C5E5CF}-4.6 & 25.51 & \cellcolor[HTML]{BEE3CA}-5.2 & 25.23 & \cellcolor[HTML]{B2DEBF}-6.3& 24.89 & \cellcolor[HTML]{A3D7B2}-7.6& 24.80 & \cellcolor[HTML]{9FD6AE}-8.0& 24.63 & \cellcolor[HTML]{97D3A7}-8.7\\
&& ResLoad& 25.78 & \cellcolor[HTML]{CBE8D4}-4.1 & 25.66 & \cellcolor[HTML]{C5E5D0}-4.6 & 25.48 & \cellcolor[HTML]{BDE2C9}-5.3 & 24.87 & \cellcolor[HTML]{A2D7B1}-7.7& 24.38 & \cellcolor[HTML]{8BCE9D}-9.7& 24.05 & \cellcolor[HTML]{7CC890}-11.1 & 23.92 & \cellcolor[HTML]{75C58B}-11.6 \\
&& Load, RES& 25.23 & \cellcolor[HTML]{B2DEBF}-6.3 & 25.06 & \cellcolor[HTML]{AADBB8}-7.0 & 24.85 & \cellcolor[HTML]{A1D7B0}-7.8 & 24.31 & \cellcolor[HTML]{88CD9B}-10.0 & 23.88 & \cellcolor[HTML]{74C489}-11.8 & 23.74 & \cellcolor[HTML]{6DC283}-12.4 & 23.52 & \cellcolor[HTML]{63BE7B}-13.3 \\
\multirow{-28}{*}{\rotatebox{90}{Expert}} & \multirow{-7}{*}{\rotatebox{90}{QR}}& Load, Solar, Wind & 25.44 & \cellcolor[HTML]{BCE2C7}-5.4 & 25.27 & \cellcolor[HTML]{B4DEC1}-6.1 & 25.09 & \cellcolor[HTML]{ACDBBA}-6.8 & 24.59 & \cellcolor[HTML]{95D2A6}-8.9& 24.21 & \cellcolor[HTML]{83CB97}-10.4 & 23.92 & \cellcolor[HTML]{76C58B}-11.6 & 23.75 & \cellcolor[HTML]{6DC284}-12.3 \\

\end{tabular}
}
\end{table*}

\newpage

\begin{table*}[htb!]
\caption{{Root mean square error and percentage improvement over the benchmark (\%chng) of HLM-based forecasting models. The reported percentage change refers to the improvement over the model without any probabilistic inputs. The green color indicates the best performing models.}}
\label{tab:app2}
\scalebox{0.7}{
\begin{tabular}{rrr|cr|cr|cr|cr|cr|cr|cr}

\noalign{\smallskip}
& \multicolumn{1}{l}{--} & -- & \multicolumn{14}{c}{24.17} \\
 \multicolumn{1}{l}{} & \multicolumn{1}{l}{} && \multicolumn{2}{c|}{$\mathcal{T}_5$} & \multicolumn{2}{c|}{$\mathcal{T}_7$} & \multicolumn{2}{c|}{$\mathcal{T}_{11}$} & \multicolumn{2}{c|}{$\mathcal{T}_{21}$} & \multicolumn{2}{c|}{$\mathcal{T}_{51}$} & \multicolumn{2}{c|}{$\mathcal{T}_{101}$} & \multicolumn{2}{c}{$\mathcal{T}_{201}$} \\
 &&& \multicolumn{1}{c}{\rotatebox{90}{RMSE}} & \multicolumn{1}{c|}{\rotatebox{90}{\%chng}} & \multicolumn{1}{c}{\rotatebox{90}{RMSE}} & \multicolumn{1}{c|}{\rotatebox{90}{\%chng}} & \multicolumn{1}{c}{\rotatebox{90}{RMSE}} & \multicolumn{1}{c|}{\rotatebox{90}{\%chng}} & \multicolumn{1}{c}{\rotatebox{90}{RMSE}} & \multicolumn{1}{c|}{\rotatebox{90}{\%chng}} & \multicolumn{1}{c}{\rotatebox{90}{RMSE}} & \multicolumn{1}{c|}{\rotatebox{90}{\%chng}} & \multicolumn{1}{c}{\rotatebox{90}{RMSE}} & \multicolumn{1}{c|}{\rotatebox{90}{\%chng}} & \multicolumn{1}{c}{\rotatebox{90}{RMSE}} & \multicolumn{1}{c}{\rotatebox{90}{\%chng}} \\
 \cline{2-17}
 \noalign{\smallskip}
&& Load& 24.13 & \cellcolor[HTML]{F2F8F7}-0.2 & 24.17 & \cellcolor[HTML]{F6F9FA}0.0& 24.17 & \cellcolor[HTML]{F6F9FA}0.0& 24.13 & \cellcolor[HTML]{F2F8F7}-0.2& 24.13 & \cellcolor[HTML]{F2F8F6}-0.2& 24.10 & \cellcolor[HTML]{EFF7F4}-0.3& 24.08 & \cellcolor[HTML]{EDF6F2}-0.4\\
&& Solar & 24.17 & \cellcolor[HTML]{F6F9FA}0.0& 24.17 & \cellcolor[HTML]{F6F9FA}0.0& 24.15 & \cellcolor[HTML]{F4F9F8}-0.1 & 24.15 & \cellcolor[HTML]{F4F8F8}-0.1& 24.17 & \cellcolor[HTML]{F6F9FA}0.0& 24.15 & \cellcolor[HTML]{F4F8F8}-0.1& 24.15 & \cellcolor[HTML]{F4F9F8}-0.1\\
&& Wind& 24.20 & \cellcolor[HTML]{F9FAFC}0.1& 24.21 & \cellcolor[HTML]{FAFBFD}0.1& 24.18 & \cellcolor[HTML]{F7FAFB}0.0& 24.17 & \cellcolor[HTML]{F6F9FA}0.0& 24.19 & \cellcolor[HTML]{F8FAFB}0.0& 24.16 & \cellcolor[HTML]{F5F9F9}0.0& 24.13 & \cellcolor[HTML]{F2F7F6}-0.2\\
&& RES& 24.13 & \cellcolor[HTML]{F2F8F7}-0.2 & 24.09 & \cellcolor[HTML]{EEF6F3}-0.3 & 24.08 & \cellcolor[HTML]{EDF6F2}-0.4 & 24.07 & \cellcolor[HTML]{ECF5F1}-0.4& 24.06 & \cellcolor[HTML]{EBF5F1}-0.5& 24.07 & \cellcolor[HTML]{ECF5F1}-0.4& 24.08 & \cellcolor[HTML]{EDF5F2}-0.4\\
&& ResLoad& 24.03 & \cellcolor[HTML]{E8F4EE}-0.6 & 23.99 & \cellcolor[HTML]{E4F2EA}-0.8 & 23.94 & \cellcolor[HTML]{DFF0E6}-1.0 & 23.95 & \cellcolor[HTML]{E0F0E7}-0.9& 23.94 & \cellcolor[HTML]{DFF0E5}-1.0& 23.95 & \cellcolor[HTML]{DFF0E6}-1.0& 23.93 & \cellcolor[HTML]{DEEFE5}-1.0\\
&& Load, RES& 24.08 & \cellcolor[HTML]{EDF6F2}-0.4 & 24.07 & \cellcolor[HTML]{ECF5F1}-0.5 & 24.04 & \cellcolor[HTML]{E9F4EE}-0.6 & 23.99 & \cellcolor[HTML]{E4F2EA}-0.8& 24.04 & \cellcolor[HTML]{E9F4EF}-0.5& 23.99 & \cellcolor[HTML]{E4F2EA}-0.8& 23.98 & \cellcolor[HTML]{E3F1E9}-0.8\\
& \multirow{-7}{*}{\rotatebox{90}{ReLU}} & Load, Solar, Wind & 24.14 & \cellcolor[HTML]{F3F8F8}-0.1 & 24.15 & \cellcolor[HTML]{F4F8F8}-0.1 & 24.10 & \cellcolor[HTML]{EFF7F4}-0.3 & 24.09 & \cellcolor[HTML]{EEF6F3}-0.4& 24.04 & \cellcolor[HTML]{E9F4EE}-0.6& 24.00 & \cellcolor[HTML]{E5F2EB}-0.7& 24.00 & \cellcolor[HTML]{E5F2EB}-0.7\\

\cline{2-17}
\noalign{\smallskip}
&& Load& 24.14 & \cellcolor[HTML]{F4F8F8}-0.1 & 24.18 & \cellcolor[HTML]{F7FAFA}0.0& 24.17 & \cellcolor[HTML]{F6F9FA}0.0& 24.16 & \cellcolor[HTML]{F5F9F9}-0.1& 24.16 & \cellcolor[HTML]{F5F9F9}-0.1& 24.16 & \cellcolor[HTML]{F6F9F9}0.0& 24.17 & \cellcolor[HTML]{F6F9FA}0.0\\
&& Solar & 24.20 & \cellcolor[HTML]{F9FAFC}0.1& 24.22 & \cellcolor[HTML]{FCFCFF}0.2& 24.18 & \cellcolor[HTML]{F7FAFB}0.0& 24.17 & \cellcolor[HTML]{F6F9FA}0.0& 24.21 & \cellcolor[HTML]{FAFBFD}0.1& 24.18 & \cellcolor[HTML]{F7FAFA}0.0& 24.17 & \cellcolor[HTML]{F6F9F9}0.0\\
&& Wind& 24.19 & \cellcolor[HTML]{F8FAFC}0.1& 24.19 & \cellcolor[HTML]{F8FAFC}0.1& 24.18 & \cellcolor[HTML]{F7FAFB}0.0& 24.19 & \cellcolor[HTML]{F8FAFB}0.1& 24.17 & \cellcolor[HTML]{F6F9FA}0.0& 24.17 & \cellcolor[HTML]{F7F9FA}0.0& 24.15 & \cellcolor[HTML]{F4F9F8}-0.1\\
&& RES& 24.17 & \cellcolor[HTML]{F6F9FA}0.0& 24.16 & \cellcolor[HTML]{F5F9F9}-0.1 & 24.17 & \cellcolor[HTML]{F6F9FA}0.0& 24.16 & \cellcolor[HTML]{F5F9F9}-0.1& 24.16 & \cellcolor[HTML]{F5F9F9}-0.1& 24.15 & \cellcolor[HTML]{F4F9F8}-0.1& 24.16 & \cellcolor[HTML]{F5F9F9}-0.1\\
&& ResLoad& 24.15 & \cellcolor[HTML]{F4F8F8}-0.1 & 24.16 & \cellcolor[HTML]{F5F9F9}0.0& 24.14 & \cellcolor[HTML]{F3F8F7}-0.1 & 24.16 & \cellcolor[HTML]{F5F9F9}-0.1& 24.16 & \cellcolor[HTML]{F5F9F9}-0.1& 24.11 & \cellcolor[HTML]{F0F7F4}-0.3& 24.15 & \cellcolor[HTML]{F5F9F8}-0.1\\
&& Load, RES& 24.13 & \cellcolor[HTML]{F2F8F6}-0.2 & 24.14 & \cellcolor[HTML]{F3F8F7}-0.1 & 24.14 & \cellcolor[HTML]{F3F8F7}-0.2 & 24.14 & \cellcolor[HTML]{F3F8F7}-0.1& 24.15 & \cellcolor[HTML]{F4F9F8}-0.1& 24.13 & \cellcolor[HTML]{F3F8F7}-0.2& 24.13 & \cellcolor[HTML]{F2F8F6}-0.2\\
& \multirow{-7}{*}{\rotatebox{90}{HS}}& Load, Solar, Wind & 24.16 & \cellcolor[HTML]{F5F9F9}0.0& 24.14 & \cellcolor[HTML]{F3F8F8}-0.1 & 24.17 & \cellcolor[HTML]{F6F9FA}0.0& 24.17 & \cellcolor[HTML]{F6F9FA}0.0& 24.13 & \cellcolor[HTML]{F2F8F7}-0.2& 24.13 & \cellcolor[HTML]{F2F8F7}-0.2& 24.14 & \cellcolor[HTML]{F3F8F7}-0.1\\

 \cline{2-17}
\noalign{\smallskip}

& & Load & 24.18 & \cellcolor[HTML]{F7FAFA}0.0 & 24.2 & \cellcolor[HTML]{F7F9FA}0.0 & 24.18 & \cellcolor[HTML]{F7FAFA}0.0 & 24.2 & \cellcolor[HTML]{F5F9F9}-0.1 & 24.17 & \cellcolor[HTML]{F6F9FA}0.0 & 24.2 & \cellcolor[HTML]{F5F9F9}0.0 & 24.15 & \cellcolor[HTML]{F4F8F8}-0.1 \\
 & & Solar & 24.19 & \cellcolor[HTML]{F8FAFC}0.1 & 24.2 & \cellcolor[HTML]{FAFBFD}0.1 & 24.20 & \cellcolor[HTML]{F9FBFD}0.1 & 24.2 & \cellcolor[HTML]{FAFBFD}0.1 & 24.19 & \cellcolor[HTML]{F8FAFC}0.1 & 24.2 & \cellcolor[HTML]{F7FAFA}0.0 & 24.20 & \cellcolor[HTML]{F9FAFC}0.1 \\
 & & Wind & 24.15 & \cellcolor[HTML]{F4F9F8}-0.1 & 24.2 & \cellcolor[HTML]{F7FAFB}0.0 & 24.18 & \cellcolor[HTML]{F7FAFB}0.0 & 24.2 & \cellcolor[HTML]{F7FAFA}0.0 & 24.18 & \cellcolor[HTML]{F7FAFB}0.0 & 24.2 & \cellcolor[HTML]{F5F9F9}-0.1 & 24.14 & \cellcolor[HTML]{F3F8F8}-0.1 \\
 & & RES & 24.16 & \cellcolor[HTML]{F5F9F9}-0.1 & 24.2 & \cellcolor[HTML]{F6F9FA}0.0 & 24.17 & \cellcolor[HTML]{F7F9FA}0.0 & 24.2 & \cellcolor[HTML]{F6F9FA}0.0 & 24.16 & \cellcolor[HTML]{F5F9F9}-0.1 & 24.2 & \cellcolor[HTML]{F4F9F8}-0.1 & 24.16 & \cellcolor[HTML]{F5F9F9}-0.1 \\
 & & ResLoad & 24.17 & \cellcolor[HTML]{F6F9FA}0.0 & 24.2 & \cellcolor[HTML]{F5F9F8}-0.1 & 24.15 & \cellcolor[HTML]{F5F9F8}-0.1 & 24.2 & \cellcolor[HTML]{F5F9F9}-0.1 & 24.12 & \cellcolor[HTML]{F1F7F5}-0.2 & 24.2 & \cellcolor[HTML]{F6F9FA}0.0 & 24.15 & \cellcolor[HTML]{F4F9F8}-0.1 \\
 & & Load, RES & 24.17 & \cellcolor[HTML]{F6F9FA}0.0 & 24.2 & \cellcolor[HTML]{F6F9FA}0.0 & 24.17 & \cellcolor[HTML]{F7FAFA}0.0 & 24.1 & \cellcolor[HTML]{F1F7F5}-0.2 & 24.16 & \cellcolor[HTML]{F6F9F9}0.0 & 24.1 & \cellcolor[HTML]{F4F8F8}-0.1 & 24.14 & \cellcolor[HTML]{F3F8F8}-0.1 \\
 & \multirow{-7}{*}{\rotatebox{90}{CP}} & Load, Solar, Wind & 24.17 & \cellcolor[HTML]{F6F9FA}0.0 & 24.2 & \cellcolor[HTML]{F8FAFB}0.0 & 24.16 & \cellcolor[HTML]{F5F9F9}-0.1 & 24.2 & \cellcolor[HTML]{F6F9FA}0.0 & 24.17 & \cellcolor[HTML]{F6F9FA}0.0 & 24.2 & \cellcolor[HTML]{F4F9F8}-0.1 & 24.14 & \cellcolor[HTML]{F3F8F8}-0.1 \\
 
\cline{2-17}
\noalign{\smallskip}
&& Load& 23.97 & \cellcolor[HTML]{E2F1E9}-0.8 & 23.98 & \cellcolor[HTML]{E3F2E9}-0.8 & 23.98 & \cellcolor[HTML]{E3F2EA}-0.8 & 23.94 & \cellcolor[HTML]{DFF0E6}-1.0& 23.93 & \cellcolor[HTML]{DEEFE5}-1.0& 23.94 & \cellcolor[HTML]{DFF0E6}-1.0& 23.92 & \cellcolor[HTML]{DDEFE4}-1.1\\
&& Solar & 24.18 & \cellcolor[HTML]{F7FAFA}0.0& 24.17 & \cellcolor[HTML]{F6F9FA}0.0& 24.18 & \cellcolor[HTML]{F7FAFB}0.0& 24.16 & \cellcolor[HTML]{F5F9F9}0.0& 24.17 & \cellcolor[HTML]{F6F9FA}0.0& 24.17 & \cellcolor[HTML]{F6F9FA}0.0& 24.15 & \cellcolor[HTML]{F4F9F8}-0.1\\
&& Wind& 23.94 & \cellcolor[HTML]{DFF0E6}-1.0 & 23.95 & \cellcolor[HTML]{E0F0E7}-0.9 & 23.93 & \cellcolor[HTML]{DEF0E5}-1.0 & 23.87 & \cellcolor[HTML]{D7EDDF}-1.3& 23.76 & \cellcolor[HTML]{CDE9D6}-1.7& 23.72 & \cellcolor[HTML]{C8E7D2}-1.9& 23.62 & \cellcolor[HTML]{BEE3CA}-2.3\\
&& RES& 23.63 & \cellcolor[HTML]{BFE3CA}-2.3 & 23.66 & \cellcolor[HTML]{C2E4CD}-2.1 & 23.66 & \cellcolor[HTML]{C2E4CD}-2.1 & 23.59 & \cellcolor[HTML]{BBE1C7}-2.4& 23.42 & \cellcolor[HTML]{A9DAB8}-3.2& 23.38 & \cellcolor[HTML]{A5D8B4}-3.4& 23.26 & \cellcolor[HTML]{98D3A9}-3.9\\
&& ResLoad& 23.61 & \cellcolor[HTML]{BDE2C9}-2.4 & 23.56 & \cellcolor[HTML]{B7E0C4}-2.6 & 23.52 & \cellcolor[HTML]{B3DEC0}-2.8 & 23.28 & \cellcolor[HTML]{9BD4AB}-3.8& 23.06 & \cellcolor[HTML]{83CB97}-4.7& 22.84 & \cellcolor[HTML]{6CC183}-5.7& 22.75 & \cellcolor[HTML]{63BE7B}-6.1\\
&& Load, RES& 23.39 & \cellcolor[HTML]{A6D9B5}-3.3 & 23.40 & \cellcolor[HTML]{A7D9B6}-3.3 & 23.40 & \cellcolor[HTML]{A8D9B6}-3.2 & 23.22 & \cellcolor[HTML]{94D2A5}-4.0& 23.08 & \cellcolor[HTML]{86CC99}-4.6& 22.96 & \cellcolor[HTML]{79C68E}-5.2& 22.83 & \cellcolor[HTML]{6BC182}-5.7\\
\multirow{-28}{*}{\rotatebox{90}{HLM}} & \multirow{-7}{*}{\rotatebox{90}{QR}}& Load, Solar, Wind & 23.68 & \cellcolor[HTML]{C4E5CF}-2.1 & 23.68 & \cellcolor[HTML]{C4E5CE}-2.1 & 23.63 & \cellcolor[HTML]{BFE3CB}-2.3 & 23.52 & \cellcolor[HTML]{B4DFC1}-2.7& 23.29 & \cellcolor[HTML]{9CD5AC}-3.7& 23.20 & \cellcolor[HTML]{92D1A4}-4.1& 23.10 & \cellcolor[HTML]{88CD9B}-4.5

\end{tabular}
}
\end{table*}

\newpage 

\section{Mean absolute error (MAE) for all considered models}

\begin{table*}[htb!]
\caption{{Mean absolute error and percentage improvement over the benchmark (\%chng) of Expert-based forecasting models. The reported percentage change refers to the improvement over the model without any probabilistic inputs. The green color indicates the best performing models.}}
\label{tab:app3}
\scalebox{0.7}{
\begin{tabular}{rrr|cr|cr|cr|cr|cr|cr|cr}
 & \multicolumn{1}{l}{--} & -- & \multicolumn{14}{c}{14.63} \\
 \multicolumn{1}{l}{} & \multicolumn{1}{l}{} && \multicolumn{2}{c|}{$\mathcal{T}_5$} & \multicolumn{2}{c|}{$\mathcal{T}_7$} & \multicolumn{2}{c|}{$\mathcal{T}_{11}$} & \multicolumn{2}{c|}{$\mathcal{T}_{21}$} & \multicolumn{2}{c|}{$\mathcal{T}_{51}$} & \multicolumn{2}{c|}{$\mathcal{T}_{101}$} & \multicolumn{2}{c}{$\mathcal{T}_{201}$} \\
 &&& \multicolumn{1}{c}{\rotatebox{90}{MAE}} & \multicolumn{1}{c|}{\rotatebox{90}{\%chng}} & \multicolumn{1}{c}{\rotatebox{90}{MAE}} & \multicolumn{1}{c|}{\rotatebox{90}{\%chng}} & \multicolumn{1}{c}{\rotatebox{90}{MAE}} & \multicolumn{1}{c|}{\rotatebox{90}{\%chng}} & \multicolumn{1}{c}{\rotatebox{90}{MAE}} & \multicolumn{1}{c|}{\rotatebox{90}{\%chng}} & \multicolumn{1}{c}{\rotatebox{90}{MAE}} & \multicolumn{1}{c|}{\rotatebox{90}{\%chng}} & \multicolumn{1}{c}{\rotatebox{90}{MAE}} & \multicolumn{1}{c|}{\rotatebox{90}{\%chng}} & \multicolumn{1}{c}{\rotatebox{90}{MAE}} & \multicolumn{1}{c}{\rotatebox{90}{\%chng}} \\
 \cline{2-17}
 \noalign{\smallskip}
 & & Load & 14.60 & \cellcolor[HTML]{F5F9F9}-0.2 & 14.67 & \cellcolor[HTML]{FBFBFE}0.3 & 14.65 & \cellcolor[HTML]{F9FBFD}0.2 & 14.59 & \cellcolor[HTML]{F4F8F8}-0.3 & 14.52 & \cellcolor[HTML]{EFF6F3}-0.7 & 14.50 & \cellcolor[HTML]{ECF5F1}-0.9 & 14.49 & \cellcolor[HTML]{ECF5F1}-0.9 \\
 & & Solar & 14.59 & \cellcolor[HTML]{F5F9F9}-0.2 & 14.62 & \cellcolor[HTML]{F7F9FA}-0.1 & 14.61 & \cellcolor[HTML]{F6F9FA}-0.1 & 14.59 & \cellcolor[HTML]{F4F9F8}-0.2 & 14.59 & \cellcolor[HTML]{F5F9F9}-0.2 & 14.59 & \cellcolor[HTML]{F4F9F8}-0.2 & 14.60 & \cellcolor[HTML]{F5F9F9}-0.2 \\
 & & Wind & 14.58 & \cellcolor[HTML]{F3F8F8}-0.3 & 14.59 & \cellcolor[HTML]{F5F9F9}-0.2 & 14.59 & \cellcolor[HTML]{F4F9F8}-0.2 & 14.55 & \cellcolor[HTML]{F1F7F5}-0.5 & 14.51 & \cellcolor[HTML]{EDF6F2}-0.8 & 14.46 & \cellcolor[HTML]{E9F4EF}-1.1 & 14.47 & \cellcolor[HTML]{E9F4EF}-1.1 \\
 & & RES & 14.54 & \cellcolor[HTML]{F0F7F5}-0.6 & 14.50 & \cellcolor[HTML]{ECF5F1}-0.9 & 14.50 & \cellcolor[HTML]{ECF5F1}-0.9 & 14.44 & \cellcolor[HTML]{E7F3ED}-1.3 & 14.34 & \cellcolor[HTML]{DEEFE5}-2.0 & 14.33 & \cellcolor[HTML]{DEEFE5}-2.0 & 14.32 & \cellcolor[HTML]{DCEFE3}-2.1 \\
 & & ResLoad & 14.34 & \cellcolor[HTML]{DEF0E5}-2.0 & 14.27 & \cellcolor[HTML]{D8EDE0}-2.5 & 14.15 & \cellcolor[HTML]{CDE9D7}-3.3 & 14.05 & \cellcolor[HTML]{C4E5CF}-4.0 & 13.94 & \cellcolor[HTML]{BAE1C6}-4.8 & 13.92 & \cellcolor[HTML]{B8E0C5}-4.9 & 13.91 & \cellcolor[HTML]{B7E0C4}-5.0 \\
 & & Load, RES & 14.51 & \cellcolor[HTML]{EDF6F2}-0.8 & 14.54 & \cellcolor[HTML]{F0F7F4}-0.6 & 14.48 & \cellcolor[HTML]{EBF5F0}-1.0 & 14.36 & \cellcolor[HTML]{E0F0E6}-1.9 & 14.14 & \cellcolor[HTML]{CCE8D6}-3.4 & 14.07 & \cellcolor[HTML]{C6E6D0}-3.9 & 13.98 & \cellcolor[HTML]{BEE3CA}-4.5 \\
 & \multirow{-7}{*}{\rotatebox{90}{ReLU}} & Load, Solar, Wind & 14.52 & \cellcolor[HTML]{EEF6F3}-0.7 & 14.61 & \cellcolor[HTML]{F6F9FA}-0.1 & 14.56 & \cellcolor[HTML]{F2F8F6}-0.4 & 14.41 & \cellcolor[HTML]{E5F2EB}-1.5 & 14.28 & \cellcolor[HTML]{D9EDE1}-2.4 & 14.12 & \cellcolor[HTML]{CAE7D4}-3.5 & 14.05 & \cellcolor[HTML]{C4E5CE}-4.0 \\

\cline{2-17}
\noalign{\smallskip}
 
& & Load & 14.57 & \cellcolor[HTML]{F3F8F7}-0.4 & 14.58 & \cellcolor[HTML]{F3F8F7}-0.3 & 14.58 & \cellcolor[HTML]{F4F8F8}-0.3 & 14.57 & \cellcolor[HTML]{F3F8F7}-0.4 & 14.55 & \cellcolor[HTML]{F1F7F6}-0.5 & 14.56 & \cellcolor[HTML]{F2F8F6}-0.4 & 14.58 & \cellcolor[HTML]{F3F8F7}-0.3 \\
 & & Solar & 14.65 & \cellcolor[HTML]{FAFBFD}0.2 & 14.66 & \cellcolor[HTML]{FAFBFD}0.2 & 14.67 & \cellcolor[HTML]{FBFBFE}0.3 & 14.67 & \cellcolor[HTML]{FCFCFF}0.3 & 14.67 & \cellcolor[HTML]{FBFBFE}0.3 & 14.66 & \cellcolor[HTML]{FBFBFE}0.2 & 14.66 & \cellcolor[HTML]{FBFBFE}0.2 \\
 & & Wind & 14.61 & \cellcolor[HTML]{F6F9FA}-0.1 & 14.61 & \cellcolor[HTML]{F6F9FA}-0.1 & 14.62 & \cellcolor[HTML]{F7FAFA}-0.1 & 14.61 & \cellcolor[HTML]{F6F9FA}-0.1 & 14.59 & \cellcolor[HTML]{F4F9F8}-0.2 & 14.59 & \cellcolor[HTML]{F4F9F8}-0.2 & 14.60 & \cellcolor[HTML]{F5F9F9}-0.2 \\
 & & RES & 14.61 & \cellcolor[HTML]{F6F9FA}-0.1 & 14.60 & \cellcolor[HTML]{F5F9F9}-0.2 & 14.60 & \cellcolor[HTML]{F5F9F9}-0.2 & 14.59 & \cellcolor[HTML]{F4F9F8}-0.3 & 14.58 & \cellcolor[HTML]{F3F8F7}-0.3 & 14.56 & \cellcolor[HTML]{F2F8F6}-0.4 & 14.55 & \cellcolor[HTML]{F1F7F5}-0.5 \\
 & & ResLoad & 14.61 & \cellcolor[HTML]{F6F9FA}-0.1 & 14.60 & \cellcolor[HTML]{F5F9F9}-0.2 & 14.59 & \cellcolor[HTML]{F4F9F8}-0.2 & 14.59 & \cellcolor[HTML]{F4F9F8}-0.2 & 14.60 & \cellcolor[HTML]{F5F9F9}-0.2 & 14.57 & \cellcolor[HTML]{F3F8F7}-0.4 & 14.58 & \cellcolor[HTML]{F3F8F7}-0.3 \\
 & & Load, RES & 14.54 & \cellcolor[HTML]{F0F7F5}-0.6 & 14.54 & \cellcolor[HTML]{F0F7F4}-0.6 & 14.56 & \cellcolor[HTML]{F2F7F6}-0.5 & 14.54 & \cellcolor[HTML]{F0F7F4}-0.6 & 14.53 & \cellcolor[HTML]{EFF6F4}-0.7 & 14.52 & \cellcolor[HTML]{EEF6F3}-0.7 & 14.51 & \cellcolor[HTML]{EDF6F2}-0.8 \\
 & \multirow{-7}{*}{\rotatebox{90}{HS}} & Load, Solar, Wind & 14.59 & \cellcolor[HTML]{F4F8F8}-0.3 & 14.58 & \cellcolor[HTML]{F3F8F7}-0.3 & 14.58 & \cellcolor[HTML]{F3F8F7}-0.3 & 14.56 & \cellcolor[HTML]{F2F8F6}-0.4 & 14.53 & \cellcolor[HTML]{EFF6F4}-0.7 & 14.51 & \cellcolor[HTML]{EEF6F3}-0.8 & 14.50 & \cellcolor[HTML]{ECF5F1}-0.9 \\
 
\cline{2-17}
\noalign{\smallskip}

 & & Load & 14.57 & \cellcolor[HTML]{F3F8F7}-0.4 & 14.6 & \cellcolor[HTML]{F4F8F8}-0.3 & 14.58 & \cellcolor[HTML]{F3F8F7}-0.3 & 14.6 & \cellcolor[HTML]{F4F9F8}-0.3 & 14.58 & \cellcolor[HTML]{F3F8F7}-0.3 & 14.6 & \cellcolor[HTML]{F2F8F6}-0.4 & 14.58 & \cellcolor[HTML]{F3F8F7}-0.3 \\
 & & Solar & 14.64 & \cellcolor[HTML]{F8FAFC}0.1 & 14.6 & \cellcolor[HTML]{F8FAFC}0.1 & 14.64 & \cellcolor[HTML]{F9FAFC}0.1 & 14.6 & \cellcolor[HTML]{F9FAFC}0.1 & 14.64 & \cellcolor[HTML]{F8FAFC}0.1 & 14.6 & \cellcolor[HTML]{F9FBFD}0.1 & 14.64 & \cellcolor[HTML]{F9FBFC}0.1 \\
 & & Wind & 14.64 & \cellcolor[HTML]{F9FAFC}0.1 & 14.6 & \cellcolor[HTML]{F9FAFC}0.1 & 14.65 & \cellcolor[HTML]{F9FBFD}0.1 & 14.6 & \cellcolor[HTML]{F9FBFC}0.1 & 14.65 & \cellcolor[HTML]{FAFBFD}0.2 & 14.6 & \cellcolor[HTML]{F7FAFB}0.0 & 14.61 & \cellcolor[HTML]{F6F9FA}-0.1 \\
 & & RES & 14.65 & \cellcolor[HTML]{FAFBFD}0.2 & 14.7 & \cellcolor[HTML]{FBFBFE}0.3 & 14.66 & \cellcolor[HTML]{FAFBFE}0.2 & 14.6 & \cellcolor[HTML]{F9FBFD}0.1 & 14.64 & \cellcolor[HTML]{F9FAFC}0.1 & 14.6 & \cellcolor[HTML]{F7FAFB}0.0 & 14.61 & \cellcolor[HTML]{F6F9FA}-0.1 \\
 & & ResLoad & 14.56 & \cellcolor[HTML]{F2F8F6}-0.4 & 14.6 & \cellcolor[HTML]{F3F8F7}-0.4 & 14.57 & \cellcolor[HTML]{F3F8F7}-0.4 & 14.6 & \cellcolor[HTML]{F2F7F6}-0.5 & 14.55 & \cellcolor[HTML]{F1F7F5}-0.5 & 14.5 & \cellcolor[HTML]{EEF6F3}-0.7 & 14.50 & \cellcolor[HTML]{EDF6F2}-0.8 \\
 & & Load, RES & 14.59 & \cellcolor[HTML]{F5F9F9}-0.2 & 14.6 & \cellcolor[HTML]{F4F8F8}-0.3 & 14.59 & \cellcolor[HTML]{F5F9F9}-0.2 & 14.6 & \cellcolor[HTML]{F4F8F8}-0.3 & 14.57 & \cellcolor[HTML]{F3F8F7}-0.4 & 14.5 & \cellcolor[HTML]{F0F7F5}-0.5 & 14.53 & \cellcolor[HTML]{EFF6F3}-0.7 \\
 & \multirow{-7}{*}{\rotatebox{90}{CP}} & Load, Solar, Wind & 14.60 & \cellcolor[HTML]{F5F9F9}-0.2 & 14.6 & \cellcolor[HTML]{F3F8F7}-0.3 & 14.58 & \cellcolor[HTML]{F4F8F8}-0.3 & 14.6 & \cellcolor[HTML]{F3F8F7}-0.3 & 14.55 & \cellcolor[HTML]{F1F7F5}-0.5 & 14.5 & \cellcolor[HTML]{F0F7F4}-0.6 & 14.53 & \cellcolor[HTML]{EFF6F3}-0.7 \\

 \cline{2-17}
\noalign{\smallskip}

 & & Load & 14.43 & \cellcolor[HTML]{E7F3ED}-1.3 & 14.41 & \cellcolor[HTML]{E5F2EB}-1.5 & 14.40 & \cellcolor[HTML]{E4F2EA}-1.5 & 14.33 & \cellcolor[HTML]{DDEFE4}-2.0 & 14.30 & \cellcolor[HTML]{DBEEE3}-2.2 & 14.28 & \cellcolor[HTML]{D9EDE1}-2.4 & 14.28 & \cellcolor[HTML]{D9EDE1}-2.4 \\
 & & Solar & 14.61 & \cellcolor[HTML]{F6F9FA}-0.1 & 14.61 & \cellcolor[HTML]{F6F9FA}-0.1 & 14.60 & \cellcolor[HTML]{F5F9F9}-0.2 & 14.59 & \cellcolor[HTML]{F4F9F8}-0.3 & 14.58 & \cellcolor[HTML]{F3F8F8}-0.3 & 14.59 & \cellcolor[HTML]{F5F9F8}-0.2 & 14.58 & \cellcolor[HTML]{F4F8F8}-0.3 \\
 & & Wind & 14.36 & \cellcolor[HTML]{E0F0E7}-1.8 & 14.33 & \cellcolor[HTML]{DDEFE4}-2.0 & 14.27 & \cellcolor[HTML]{D8EDE0}-2.5 & 14.22 & \cellcolor[HTML]{D4EBDC}-2.8 & 14.11 & \cellcolor[HTML]{CAE7D3}-3.6 & 14.07 & \cellcolor[HTML]{C5E6D0}-3.9 & 13.99 & \cellcolor[HTML]{BEE3CA}-4.5 \\
 & & RES & 14.20 & \cellcolor[HTML]{D1EADA}-3.0 & 14.15 & \cellcolor[HTML]{CDE9D7}-3.3 & 14.09 & \cellcolor[HTML]{C8E7D2}-3.7 & 13.97 & \cellcolor[HTML]{BDE2C9}-4.6 & 13.82 & \cellcolor[HTML]{AFDCBC}-5.7 & 13.78 & \cellcolor[HTML]{ABDBB9}-6.0 & 13.68 & \cellcolor[HTML]{A2D7B1}-6.7 \\
 & & ResLoad & 14.05 & \cellcolor[HTML]{C4E5CF}-4.0 & 13.98 & \cellcolor[HTML]{BEE3C9}-4.5 & 13.88 & \cellcolor[HTML]{B5DFC1}-5.2 & 13.60 & \cellcolor[HTML]{9AD4AB}-7.3 & 13.29 & \cellcolor[HTML]{7DC892}-9.5 & 13.15 & \cellcolor[HTML]{6FC385}-10.6 & 13.05 & \cellcolor[HTML]{66BF7D}-11.4 \\
 & & Load, RES & 13.88 & \cellcolor[HTML]{B4DFC1}-5.2 & 13.79 & \cellcolor[HTML]{ACDBBA}-5.9 & 13.69 & \cellcolor[HTML]{A3D8B2}-6.6 & 13.41 & \cellcolor[HTML]{88CD9B}-8.7 & 13.17 & \cellcolor[HTML]{72C487}-10.5 & 13.11 & \cellcolor[HTML]{6CC182}-10.9 & 13.02 & \cellcolor[HTML]{63BE7B}-11.6 \\
\multirow{-28}{*}{\rotatebox{90}{Expert}} & \multirow{-7}{*}{\rotatebox{90}{QR}} & Load, Solar, Wind & 13.84 & \cellcolor[HTML]{B1DDBE}-5.5 & 13.76 & \cellcolor[HTML]{A9DAB7}-6.1 & 13.68 & \cellcolor[HTML]{A1D7B1}-6.7 & 13.45 & \cellcolor[HTML]{8CCE9F}-8.4 & 13.28 & \cellcolor[HTML]{7CC890}-9.7 & 13.16 & \cellcolor[HTML]{70C386}-10.6 & 13.09 & \cellcolor[HTML]{69C080}-11.1 \\

\end{tabular}
}
\end{table*}

\begin{table*}[htb!]
\caption{{Mean absolute error and percentage improvement over the benchmark (\%chng) of HLM-based forecasting models. The reported percentage change refers to the improvement over the model without any probabilistic inputs. The green color indicates the best performing models.}}
\label{tab:app4}
\scalebox{0.7}{
\begin{tabular}{rrr|cr|cr|cr|cr|cr|cr|cr}

\noalign{\smallskip}
& \multicolumn{1}{l}{--} & -- & \multicolumn{14}{c}{13.13} \\
 \multicolumn{1}{l}{} & \multicolumn{1}{l}{} && \multicolumn{2}{c|}{$\mathcal{T}_5$} & \multicolumn{2}{c|}{$\mathcal{T}_7$} & \multicolumn{2}{c|}{$\mathcal{T}_{11}$} & \multicolumn{2}{c|}{$\mathcal{T}_{21}$} & \multicolumn{2}{c|}{$\mathcal{T}_{51}$} & \multicolumn{2}{c|}{$\mathcal{T}_{101}$} & \multicolumn{2}{c}{$\mathcal{T}_{201}$} \\
 &&& \multicolumn{1}{c}{\rotatebox{90}{MAE}} & \multicolumn{1}{c|}{\rotatebox{90}{\%chng}} & \multicolumn{1}{c}{\rotatebox{90}{MAE}} & \multicolumn{1}{c|}{\rotatebox{90}{\%chng}} & \multicolumn{1}{c}{\rotatebox{90}{MAE}} & \multicolumn{1}{c|}{\rotatebox{90}{\%chng}} & \multicolumn{1}{c}{\rotatebox{90}{MAE}} & \multicolumn{1}{c|}{\rotatebox{90}{\%chng}} & \multicolumn{1}{c}{\rotatebox{90}{MAE}} & \multicolumn{1}{c|}{\rotatebox{90}{\%chng}} & \multicolumn{1}{c}{\rotatebox{90}{MAE}} & \multicolumn{1}{c|}{\rotatebox{90}{\%chng}} & \multicolumn{1}{c}{\rotatebox{90}{MAE}} & \multicolumn{1}{c}{\rotatebox{90}{\%chng}} \\
 \cline{2-17}
 \noalign{\smallskip}
& & Load & 13.11 & \cellcolor[HTML]{F3F8F7}-0.2 & 13.14 & \cellcolor[HTML]{FAFBFD}0.1 & 13.15 & \cellcolor[HTML]{FCFCFF}0.1 & 13.12 & \cellcolor[HTML]{F6F9F9}-0.1 & 13.11 & \cellcolor[HTML]{F4F8F8}-0.1 & 13.10 & \cellcolor[HTML]{F0F7F5}-0.3 & 13.09 & \cellcolor[HTML]{EFF7F4}-0.3 \\
 & & Solar & 13.12 & \cellcolor[HTML]{F6F9FA}0.0 & 13.13 & \cellcolor[HTML]{F7FAFB}0.0 & 13.12 & \cellcolor[HTML]{F5F9F9}-0.1 & 13.12 & \cellcolor[HTML]{F5F9F9}-0.1 & 13.13 & \cellcolor[HTML]{F7FAFA}0.0 & 13.12 & \cellcolor[HTML]{F5F9F9}-0.1 & 13.12 & \cellcolor[HTML]{F6F9FA}-0.1 \\
 & & Wind & 13.10 & \cellcolor[HTML]{F0F7F5}-0.3 & 13.09 & \cellcolor[HTML]{F0F7F4}-0.3 & 13.09 & \cellcolor[HTML]{F0F7F4}-0.3 & 13.08 & \cellcolor[HTML]{EEF6F3}-0.3 & 13.10 & \cellcolor[HTML]{F2F8F6}-0.2 & 13.10 & \cellcolor[HTML]{F1F7F5}-0.2 & 13.08 & \cellcolor[HTML]{EEF6F3}-0.4 \\
 & & RES & 13.07 & \cellcolor[HTML]{ECF5F1}-0.4 & 13.05 & \cellcolor[HTML]{E7F3ED}-0.6 & 13.05 & \cellcolor[HTML]{E7F3ED}-0.6 & 13.06 & \cellcolor[HTML]{E9F4EF}-0.5 & 13.06 & \cellcolor[HTML]{E9F4EF}-0.5 & 13.08 & \cellcolor[HTML]{EEF6F3}-0.4 & 13.08 & \cellcolor[HTML]{ECF5F1}-0.4 \\
 & & ResLoad & 12.89 & \cellcolor[HTML]{C3E5CE}-1.9 & 12.86 & \cellcolor[HTML]{BEE2C9}-2.1 & 12.83 & \cellcolor[HTML]{B7E0C4}-2.3 & 12.86 & \cellcolor[HTML]{BDE2C9}-2.1 & 12.87 & \cellcolor[HTML]{BEE3CA}-2.0 & 12.87 & \cellcolor[HTML]{BFE3CA}-2.0 & 12.86 & \cellcolor[HTML]{BEE2C9}-2.1 \\
 & & Load, RES & 13.05 & \cellcolor[HTML]{E7F3ED}-0.6 & 13.06 & \cellcolor[HTML]{E8F4EE}-0.6 & 13.05 & \cellcolor[HTML]{E7F3EC}-0.6 & 13.03 & \cellcolor[HTML]{E3F1E9}-0.7 & 13.05 & \cellcolor[HTML]{E6F3EC}-0.6 & 13.03 & \cellcolor[HTML]{E2F1E8}-0.8 & 13.03 & \cellcolor[HTML]{E2F1E9}-0.8 \\
 & \multirow{-7}{*}{\rotatebox{90}{ReLU}} & Load, Solar, Wind & 13.07 & \cellcolor[HTML]{EBF5F0}-0.5 & 13.09 & \cellcolor[HTML]{F0F7F4}-0.3 & 13.07 & \cellcolor[HTML]{ECF5F1}-0.4 & 13.06 & \cellcolor[HTML]{E8F4EE}-0.5 & 13.04 & \cellcolor[HTML]{E5F2EB}-0.7 & 13.04 & \cellcolor[HTML]{E5F2EB}-0.7 & 13.03 & \cellcolor[HTML]{E2F1E8}-0.8 \\

\cline{2-17}
\noalign{\smallskip}
& & Load & 13.11 & \cellcolor[HTML]{F3F8F8}-0.1 & 13.12 & \cellcolor[HTML]{F6F9FA}-0.1 & 13.12 & \cellcolor[HTML]{F5F9F9}-0.1 & 13.12 & \cellcolor[HTML]{F5F9F8}-0.1 & 13.11 & \cellcolor[HTML]{F4F8F8}-0.1 & 13.12 & \cellcolor[HTML]{F5F9F9}-0.1 & 13.12 & \cellcolor[HTML]{F6F9F9}-0.1 \\
 & & Solar & 13.14 & \cellcolor[HTML]{FBFBFE}0.1 & 13.14 & \cellcolor[HTML]{FAFBFD}0.1 & 13.13 & \cellcolor[HTML]{F7FAFA}0.0 & 13.13 & \cellcolor[HTML]{F7FAFB}0.0 & 13.14 & \cellcolor[HTML]{F9FBFD}0.1 & 13.13 & \cellcolor[HTML]{F9FAFC}0.0 & 13.13 & \cellcolor[HTML]{F7FAFB}0.0 \\
 & & Wind & 13.12 & \cellcolor[HTML]{F6F9FA}0.0 & 13.12 & \cellcolor[HTML]{F5F9F9}-0.1 & 13.11 & \cellcolor[HTML]{F3F8F7}-0.2 & 13.12 & \cellcolor[HTML]{F5F9F9}-0.1 & 13.11 & \cellcolor[HTML]{F2F8F7}-0.2 & 13.11 & \cellcolor[HTML]{F3F8F7}-0.2 & 13.10 & \cellcolor[HTML]{F2F7F6}-0.2 \\
 & & RES & 13.10 & \cellcolor[HTML]{F2F8F6}-0.2 & 13.10 & \cellcolor[HTML]{F1F7F5}-0.2 & 13.11 & \cellcolor[HTML]{F3F8F7}-0.2 & 13.10 & \cellcolor[HTML]{F1F7F6}-0.2 & 13.10 & \cellcolor[HTML]{F1F7F6}-0.2 & 13.09 & \cellcolor[HTML]{EFF6F4}-0.3 & 13.10 & \cellcolor[HTML]{F1F7F5}-0.2 \\
 & & ResLoad & 13.11 & \cellcolor[HTML]{F4F8F8}-0.1 & 13.11 & \cellcolor[HTML]{F4F8F8}-0.1 & 13.12 & \cellcolor[HTML]{F5F9F9}-0.1 & 13.12 & \cellcolor[HTML]{F5F9F9}-0.1 & 13.12 & \cellcolor[HTML]{F6F9F9}-0.1 & 13.10 & \cellcolor[HTML]{F0F7F5}-0.3 & 13.12 & \cellcolor[HTML]{F5F9F9}-0.1 \\
 & & Load, RES & 13.08 & \cellcolor[HTML]{EEF6F3}-0.4 & 13.09 & \cellcolor[HTML]{EFF7F4}-0.3 & 13.09 & \cellcolor[HTML]{EFF6F4}-0.3 & 13.09 & \cellcolor[HTML]{EEF6F3}-0.3 & 13.09 & \cellcolor[HTML]{EFF7F4}-0.3 & 13.08 & \cellcolor[HTML]{EDF6F2}-0.4 & 13.07 & \cellcolor[HTML]{EBF5F0}-0.5 \\
 & \multirow{-7}{*}{\rotatebox{90}{HS}} & Load, Solar, Wind & 13.11 & \cellcolor[HTML]{F2F8F7}-0.2 & 13.10 & \cellcolor[HTML]{F1F7F5}-0.2 & 13.11 & \cellcolor[HTML]{F3F8F7}-0.2 & 13.11 & \cellcolor[HTML]{F2F8F7}-0.2 & 13.09 & \cellcolor[HTML]{F0F7F4}-0.3 & 13.09 & \cellcolor[HTML]{F0F7F5}-0.3 & 13.10 & \cellcolor[HTML]{F1F7F6}-0.2 \\

 \cline{2-17}
\noalign{\smallskip}

& & Load & 13.12 & \cellcolor[HTML]{F6F9FA}0.0 & 13.1 & \cellcolor[HTML]{F6F9F9}-0.1 & 13.12 & \cellcolor[HTML]{F7F9FA}0.0 & 13.1 & \cellcolor[HTML]{F5F9F9}-0.1 & 13.12 & \cellcolor[HTML]{F5F9F9}-0.1 & 13.1 & \cellcolor[HTML]{F5F9F9}-0.1 & 13.12 & \cellcolor[HTML]{F5F9F9}-0.1 \\
 & & Solar & 13.14 & \cellcolor[HTML]{F9FBFC}0.1 & 13.1 & \cellcolor[HTML]{FAFBFE}0.1 & 13.14 & \cellcolor[HTML]{F9FBFC}0.1 & 13.1 & \cellcolor[HTML]{FAFBFD}0.1 & 13.13 & \cellcolor[HTML]{F8FAFC}0.0 & 13.1 & \cellcolor[HTML]{F8FAFB}0.0 & 13.13 & \cellcolor[HTML]{F9FAFC}0.0 \\
 & & Wind & 13.11 & \cellcolor[HTML]{F3F8F8}-0.1 & 13.1 & \cellcolor[HTML]{F6F9F9}-0.1 & 13.12 & \cellcolor[HTML]{F5F9F8}-0.1 & 13.1 & \cellcolor[HTML]{F4F9F8}-0.1 & 13.11 & \cellcolor[HTML]{F4F8F8}-0.1 & 13.1 & \cellcolor[HTML]{F3F8F7}-0.2 & 13.10 & \cellcolor[HTML]{F1F7F5}-0.2 \\
 & & RES & 13.10 & \cellcolor[HTML]{F1F7F6}-0.2 & 13.1 & \cellcolor[HTML]{F4F8F8}-0.1 & 13.12 & \cellcolor[HTML]{F4F9F8}-0.1 & 13.1 & \cellcolor[HTML]{F2F8F6}-0.2 & 13.12 & \cellcolor[HTML]{F5F9F9}-0.1 & 13.1 & \cellcolor[HTML]{F0F7F5}-0.3 & 13.10 & \cellcolor[HTML]{F2F8F6}-0.2 \\
 & & ResLoad & 13.11 & \cellcolor[HTML]{F3F8F7}-0.2 & 13.1 & \cellcolor[HTML]{F1F7F6}-0.2 & 13.09 & \cellcolor[HTML]{F0F7F5}-0.3 & 13.1 & \cellcolor[HTML]{F1F7F6}-0.2 & 13.08 & \cellcolor[HTML]{EEF6F3}-0.3 & 13.1 & \cellcolor[HTML]{F3F8F7}-0.2 & 13.11 & \cellcolor[HTML]{F3F8F7}-0.2 \\
 & & Load, RES & 13.11 & \cellcolor[HTML]{F2F8F7}-0.2 & 13.1 & \cellcolor[HTML]{F3F8F7}-0.2 & 13.11 & \cellcolor[HTML]{F4F8F8}-0.1 & 13.1 & \cellcolor[HTML]{F0F7F5}-0.3 & 13.10 & \cellcolor[HTML]{F2F8F6}-0.2 & 13.1 & \cellcolor[HTML]{EFF7F4}-0.3 & 13.09 & \cellcolor[HTML]{F0F7F4}-0.3 \\
 & \multirow{-7}{*}{\rotatebox{90}{CP}} & Load, Solar, Wind & 13.12 & \cellcolor[HTML]{F6F9FA}-0.1 & 13.1 & \cellcolor[HTML]{F5F9F9}-0.1 & 13.11 & \cellcolor[HTML]{F3F8F7}-0.2 & 13.1 & \cellcolor[HTML]{F2F8F6}-0.2 & 13.11 & \cellcolor[HTML]{F3F8F7}-0.2 & 13.1 & \cellcolor[HTML]{F1F7F5}-0.2 & 13.10 & \cellcolor[HTML]{F1F7F6}-0.2 \\
 
\cline{2-17}
\noalign{\smallskip}
& & Load & 13.02 & \cellcolor[HTML]{DFF0E6}-0.9 & 13.01 & \cellcolor[HTML]{DEF0E5}-0.9 & 13.01 & \cellcolor[HTML]{DEF0E5}-0.9 & 13.00 & \cellcolor[HTML]{DCEFE3}-1.0 & 12.98 & \cellcolor[HTML]{D7EDDF}-1.2 & 12.99 & \cellcolor[HTML]{DAEEE2}-1.0 & 12.99 & \cellcolor[HTML]{D8EDE0}-1.1 \\
 & & Solar & 13.13 & \cellcolor[HTML]{F7FAFB}0.0 & 13.12 & \cellcolor[HTML]{F6F9FA}-0.1 & 13.13 & \cellcolor[HTML]{F7FAFB}0.0 & 13.12 & \cellcolor[HTML]{F6F9FA}-0.1 & 13.13 & \cellcolor[HTML]{F7FAFB}0.0 & 13.12 & \cellcolor[HTML]{F6F9F9}-0.1 & 13.12 & \cellcolor[HTML]{F5F9F9}-0.1 \\
 & & Wind & 13.03 & \cellcolor[HTML]{E3F2E9}-0.7 & 13.04 & \cellcolor[HTML]{E3F2EA}-0.7 & 13.02 & \cellcolor[HTML]{E1F1E8}-0.8 & 13.01 & \cellcolor[HTML]{DDEFE5}-0.9 & 12.97 & \cellcolor[HTML]{D5ECDD}-1.2 & 12.95 & \cellcolor[HTML]{D0EAD9}-1.4 & 12.92 & \cellcolor[HTML]{C9E7D3}-1.6 \\
 & & RES & 12.93 & \cellcolor[HTML]{CCE8D5}-1.6 & 12.94 & \cellcolor[HTML]{CFE9D8}-1.4 & 12.95 & \cellcolor[HTML]{D1EADA}-1.4 & 12.91 & \cellcolor[HTML]{C9E7D3}-1.7 & 12.86 & \cellcolor[HTML]{BDE2C9}-2.1 & 12.83 & \cellcolor[HTML]{B7E0C3}-2.3 & 12.79 & \cellcolor[HTML]{AEDCBB}-2.6 \\
 & & ResLoad & 12.85 & \cellcolor[HTML]{BBE1C7}-2.1 & 12.83 & \cellcolor[HTML]{B7E0C4}-2.3 & 12.82 & \cellcolor[HTML]{B5DFC1}-2.4 & 12.72 & \cellcolor[HTML]{9ED5AD}-3.2 & 12.60 & \cellcolor[HTML]{84CB97}-4.1 & 12.50 & \cellcolor[HTML]{6EC284}-4.9 & 12.45 & \cellcolor[HTML]{63BE7B}-5.3 \\
 & & Load, RES & 12.79 & \cellcolor[HTML]{AEDCBB}-2.6 & 12.79 & \cellcolor[HTML]{AEDCBC}-2.6 & 12.80 & \cellcolor[HTML]{B0DDBD}-2.6 & 12.72 & \cellcolor[HTML]{9DD5AD}-3.2 & 12.67 & \cellcolor[HTML]{92D1A4}-3.6 & 12.59 & \cellcolor[HTML]{82CA96}-4.2 & 12.55 & \cellcolor[HTML]{78C68D}-4.5 \\
\multirow{-28}{*}{\rotatebox{90}{HLM}} & \multirow{-7}{*}{\rotatebox{90}{QR}} & Load, Solar, Wind & 12.87 & \cellcolor[HTML]{C0E3CB}-2.0 & 12.87 & \cellcolor[HTML]{C0E3CB}-2.0 & 12.85 & \cellcolor[HTML]{BAE1C6}-2.2 & 12.81 & \cellcolor[HTML]{B3DEC0}-2.4 & 12.71 & \cellcolor[HTML]{9DD5AD}-3.2 & 12.66 & \cellcolor[HTML]{91D1A3}-3.6 & 12.64 & \cellcolor[HTML]{8DCF9F}-3.8 
\end{tabular}
}

\end{table*}

\end{document}